\documentclass[twocolumn, numberedappendix]{aastex63}

\usepackage{amstext}
\usepackage{amsmath}
\usepackage{amssymb}
\usepackage{graphicx}

 \usepackage{enumitem}   

\newcommand{\beq}{\begin{equation}}
\newcommand{\eeq}{\end{equation}}
\newcommand{\beqar}{\begin{align}}
\newcommand{\eeqar}{\end{align}}

\newcommand{\Md}{M_{\rm d}}

\newcommand{\Ma}{M_{\rm a}}
\newcommand{\gacc}{\gamma_{\rm a}}
\newcommand{\gloss}{\gamma_{\rm loss}}
\newcommand{\gd}{\gamma_{\rm d}}
\newcommand{\gL}{\gamma_{L_2}}

\newcommand{\dmt}{\Delta \tilde m}
\newcommand{\dme}{\Delta \tilde m_{\rm est}}

\shorttitle{}

\shortauthors{}

\begin{document}

\title{Pre-Common-Envelope Mass Loss from Coalescing Binary Systems}

\author[0000-0002-1417-8024]{Morgan MacLeod}
\affiliation{Harvard-Smithsonian Center for Astrophysics, 60 Garden Street, Cambridge, MA, 02138, USA}
\email{morgan.macleod@cfa.harvard.edu}

\author[0000-0003-4330-287X]{Abraham Loeb}
\affiliation{Harvard-Smithsonian Center for Astrophysics, 60 Garden Street, Cambridge, MA, 02138, USA}

\begin{abstract}
Binary systems undergoing unstable Roche Lobe overflow spill gas into their circumbinary environment as their orbits decay toward coalescence. In this paper, we use a suite of hydrodynamic models of coalescing binaries involving an extended donor and a more compact accretor. We focus on the period of unstable Roche Lobe overflow that ends as the accretor plunges within the envelope of the donor at the onset of a common envelope phase. During this stage, mass is removed from the donor and flung into the circumbinary environment. Across a wide range of binary mass ratios, we find that the mass expelled as the separation decreases from the Roche limit to the donor's original radius is of the order of 25\% of the accretor's mass. We study the kinematics of this ejecta and its dependencies on binary properties and find that it assembles into a toroidal circumbinary distribution. These circumbinary tori have approximately constant specific angular momentum due to momentum transport by spiral shocks launched from the orbiting binary. We show that an analytic model with these torus  properties captures many of the main features of the azimuthally-averaged profiles of our hydrodynamic simulations. Our results, in particular the simple relationship between accretor mass and expelled mass and its spatial distribution, may be useful in interpreting stellar coalescence transients like luminous red novae, and in initializing hydrodynamic simulations of the subsequent common envelope phase. 
\end{abstract}

\keywords{binaries: close, methods: numerical,  hydrodynamics}

\section{Introduction}

As observational efforts to understand the transient night sky have increased in breadth and depth, a category of optical and infrared transients, known as luminous red novae, has emerged \citep[e.g.][]{2003ApJ...582L.105S,2006MNRAS.373..733S,2013Sci...339..433I,2019A&A...630A..75P,2019ApJ...886...40J}. There is strong evidence for the association of these events with stellar coalescence episodes \citep{2011A&A...528A.114T}. The optical flare that accompanies these events is believed to arise from the heating and ejection of material from the binary system as the two stars coalesce \citep{2013Sci...339..433I,2014ApJ...786...39N,2014ApJ...788...22P,2017ApJ...850...59P,2017ApJ...835..282M,2017MNRAS.471.3200M,2018ApJ...863....5M}. 

Despite a growing body of observational evidence, a detailed theoretical understanding of how emission from these transients is generated has not yet been reached \citep[see, for example,][]{2003ApJ...582L.105S,2006MNRAS.373..733S,2013Sci...339..433I,2017ApJS..229...36G,2017MNRAS.471.3200M,2020MNRAS.492.3229H}. Lacking such a model, it is not yet possible, to infer the properties of a merging binary or common envelope phase from the transient it produces \citep[e.g.][]{2019MNRAS.490.2550I}.  The ability to make such connections is extremely desirable, however, because this would place empirical constraints on the long-uncertain physics of common envelope phases of binary star interaction. 

This paper uses hydrodynamical models for the onset of common envelope phases  to make a step toward understanding how the range of binary properties affect the ejecta that emerges when these systems coalesce \citep[following the methodology of][]{2018ApJ...863....5M,2018ApJ...868..136M,2019ApJ...877...28M,2019arXiv191205545M}. We study how ejecta mass and kinematics relate to properties of the binary system like binary mass ratio, the degree of synchronization of the donor star with the binary orbit, and donor-star structure. In a related, preceding paper, we analyzed how pre-coalescence orbital dynamics depend on these properties \citep{2019arXiv191205545M}. 

Our results are useful in understanding how the ejecta mass in observed transients might relate to the underlying binary system. We also demonstrate that circumbinary material may be a crucial component of the initial conditions of common envelope hydrodynamical simulations, which, with some exceptions \citep[e.g.][]{2008ApJ...672L..41R,2014ApJ...786...39N,2017MNRAS.464.4028I}, are often initialized at the moment of contact between the binary components \citep{2016ApJ...816L...9O,2018MNRAS.480.1898C,2019MNRAS.486.5809P}. 

In Section \ref{sec:method}, we briefly summarize the hydrodynamic method underlying our simulation models.
 In Section \ref{sec:deltam}, we define and study mass expelled from binary systems in the lead up to coalescence, when the accretor is engulfed within the envelope of the donor. 
 In Section \ref{sec:torus}, we examine the kinematics of circumbinary material at the time of coalescence, with attention to how binary parameters affect the relative quantities of bound and unbound circumbinary material. We demonstrate that a useful, though crude, approximation of the circumbinary distribution is a constant specific angular momentum hydrostatic torus, and model the dependence of torus parameters on binary system. 
In Section \ref{sec:implications}, we briefly discuss some implications of the accumulation of circumbinary material and in Section \ref{sec:conclusion} we summarize and conclude.

\section{Simulation Method and Models}\label{sec:method}

\begin{table*}[tb]
\begin{center}
\hspace{-3cm}
\begin{tabular}{cccccccccccc}
Name & $M_1$ & $m_1$ & $M_2$ & $a_0$ & $\gamma_{\rm ad}$ & $\Gamma_{\rm s}$ & $f_{\rm corot}$ &   $\Delta m(R_1)/M_1$ & $\Delta m(R_1)/M_2$ & bound frac. & unbound frac.     \\
\hline
A &1&  0.41 & 0.1 & 1.73 & 5/3 & 5/3 & 1.0 &   0.025 &  0.25 & 0.64 & 0.36  \\
B &1& 0.41 & 0.1 & 1.73 & 5/3 & 5/3 & 0.67 &  0.023 & 0.23 & 0.70 & 0.30 \\
C &1& 0.41 & 0.1 & 1.73 & 5/3 & 5/3 & 0.33 &  0.022 & 0.22 & 0.82 & 0.18  \\
D &1& 0.41 & 0.1 & 1.73 & 5/3 & 5/3 & 0.0  &  0.021 & 0.21 & 0.91 & 0.09 \\
\hline
N & 1& 0.41 & 0.01 & 1.2 & 5/3 & 5/3 & 0.0 & 0.0013 & 0.13 & 1.0 & 0.0  \\
E &1& 0.41 & 0.03 & 1.51 & 5/3 & 5/3 & 1.0  & 0.0092 & 0.31 & 0.89 & 0.11  \\
F &1& 0.41 & 0.3 & 2.06 & 5/3 & 5/3 & 1.0 &  0.086 & 0.29 & 0.65 & 0.35 \\
\hline
G &1& 0.68 & 0.1 & 1.55 & 1.35 & 1.35 & 1.0 &  0.016 & 0.16 & 0.28 & 0.72 \\
H &1& 0.68 & 0.1 & 1.55 & 1.5 & 1.35 & 1.0  & 0.014 & 0.14 & 0.36 & 0.64   \\
I &1& 0.68 & 0.1 & 1.55 & 5/3 & 1.35 & 1.0 &  0.013 & 0.13 & 0.49 & 0.51 \\
\hline
J &1& 0.68 & 0.3 & 1.75 & 1.35 & 1.35 & 1.0 & 0.053 & 0.18 &  0.20 & 0.80  \\
K &1& 0.68 & 0.3 & 1.75 & 1.5 & 1.35 & 1.0 &  0.045 & 0.15 & 0.49 & 0.51  \\
L &1& 0.68 & 0.3 & 1.75 & 5/3 & 1.35 & 1.0  & 0.046 & 0.15 & 0.79 & 0.21  \\
\hline
M & 1 & 0.41 & 0.1 & 1.55 & 5/3 & 5/3 & 1.0 & 0.022 & 0.22 & 0.60 & 0.40  \\
\hline 
\end{tabular}
\caption{Parameters of model binary systems simulated.  Columns include: donor initial mass, $M_1$, donor central mass, $r<0.3R_1$, $m_1$, accretor mass, $M_2$, initial separation, $a_0$, adiabatic index, $\gamma_{\rm ad}$, donor polytropic structural index, $\Gamma_{\rm s}$, donor fractional spin synchronization $f_{\rm corot}$, mass lost from the donor at $a=R_1$, $\Delta m (R_1)$, normalized to $M_1$ and $M_2$, and the bound and unbound fractions of $\Delta m(R_1)$, defined as described in the text. Models A-D include variations in initial spin synchronization. Models N, E, and F  are variations in the mass ratio. Models G-I and Models J-L are variations $\gamma_{\rm ad}$ with $\Gamma_{\rm s}=1.35$, and $q=0.1$ and $q=0.3$, respectively. Finally, model M is identical to A, except that it starts at a separation consistent  with models G-I.  }
\label{simtable}
\end{center}
\end{table*}

The hydrodynamic simulations on which this paper is based are performed using the {\tt Athena++} software (Stone, J. M., in preparation).\footnote{version 1.1.1 available at: \url{https://github.com/PrincetonUniversity/athena-public-version/releases/tag/v1.1.1} } 
We model mass loss from a gaseous donor star toward a point mass accretor. The donor star is modeled as a polytropic envelope surrounding a point mass core.  The calculations themselves are performed on a spherical polar computational mesh that is centered on the donor star and extends to 100 times its original radius. We, therefore, include forces associated with this non-inertial frame of reference. 

The accretor is a softened, non-absorbing point mass. The softening radius is 0.05 times the original donor-star radius, $R_1$, a length scale that was shown to achieve converged rate of orbital decay, and thus specific angular momentum imparted to ejected mass, by \citet{2018ApJ...863....5M} in their Figure 23.  The non-absorbing nature of the accretor implies that its mass is constant over the duration of the simulation. This approximation is reasonable in the limit that mass removal from the donor star is much more rapid than it can be absorbed by the accretor, which is typically the case in the extreme mass loss rate regimes (with fractions of the donor mass lost per orbital period) that are accessible via hydrodynamic simulations. However, we caution that mass accretion is possible at lower mass loss rates, and could modify the binary mass ratio and ejecta dynamics over the course of a given interaction. Given the scale-specific nature of gas cooling and accretion, we leave this to future consideration.  
Finally, in some situations, particularly where the donor and accretor have relatively similar radii (when the accretor radius is larger than 0.05 times the donor radius), the point mass approximation is not warranted, and one would want to consider the gaseous structure of the accretor as well.

A full description of this hydrodynamic method and tests is given in \citet{2018ApJ...863....5M}. This method has been applied, with minor modifications, in several follow up works \citep{2018ApJ...868..136M,2019ApJ...877...28M,2019arXiv191205545M}.
In a companion paper, \citep{2019arXiv191205545M}, we describe a parameter survey of model binary systems (we add one new simulation, model N, to this paper's analysis). These models share a system of units defined by $G=M_1 = R_1 = 1$, in which the gravitational constant and the donor's original mass, $M_1$, and radius, $R_1$, are all one.  We vary: 
\begin{enumerate}[label=(\roman*)]
\item the binary's initial mass ratio of accretor, $M_2$, to donor, $M_1$, masses, $q=M_2/M_1$;
\item the solid body rotation frequency of the donor, parameterized by degree of synchronization at the Roche limit separation, $f_{\rm corot}$;
\item the adiabatic index of the ideal gas equation of state, $\gamma_{\rm ad}$; and 
\item the structure of the donor's envelope, parameterized by the polytropic index, $\Gamma_{\rm s}$. 
\end{enumerate}
Further description of these model parameters is given in \citet{2019arXiv191205545M}. In Table \ref{simtable}, we summarize the properties and outcomes of our model binary systems.

\section{Pre-Coalescence Mass Loss}\label{sec:deltam}

Next, we examine mass lost from coalescing binaries as they transition from the Roche limit separation, where mass loss from the donor begins, to a common envelope phase -- which we define as when the two stellar cores are within the donor star's original radius.  
In \citet{2019arXiv191205545M}, we describe how mass loss from the donor star drives the evolution of the orbital angular momentum through torques applied by the gravitational force of the accretor. The key quantities, therefore, are the mass lost from the donor and the specific angular momentum it acquires through interaction with the accretor. 

\subsection{Analysis Metrics}
We measure the mass loss prior to binary coalescence, $\Delta m(a)$, by integrating the mass at radii larger than the original donor-star radius, including mass that has left the computational domain, at binary separation $a$.  We evaluate $\Delta m(a)$ in what follows at $a=R_1$, the original donor star radius. In general, however, we observe that mass ejection is still ongoing as the common envelope interaction proceeds. Thus, it is important to emphasize here that the quantities measured represent only the pre-common envelope phase, rather than the totality of the binary interaction that ensues. 

We compute the specific angular momentum with which material is lost from a binary  in which the donor mass is $\Md$ and the accretor mass is $\Ma$.\footnote{The donor mass $\Md$ decreases over the course of the simulation, and can therefore be less than the original donor mass $M_1$. The accretor mass in our simulations is constant so $\Ma = M_2$. We evaluate $\Md$ by measuring the enclosed mass within $R_1$.}   We generally report this quantity in the dimensionless form, $\gloss = l_{\rm loss}/l_{\rm bin}$, where $l_{\rm loss} = dL/dm$ and $l_{\rm bin}= \Md \Ma / M^2 \sqrt{G M a} $, in which $M=\Md + \Ma$. We will make use of a mass-loss-weighted average of $\gloss$, which we denote with brackets: $\langle \gloss  \rangle$. In \citet{2019arXiv191205545M}, we presented the following approximating formula,
\begin{align}\label{fit_gamma}
\frac{\langle \gloss \rangle - \gd}{\gL -\gd} \approx& 0.66 \left( \frac{q}{0.1} \right)^{0.08}  \left( \frac{\gamma_{\rm ad}}{5/3} \right)^{0.69}  \left( \frac{\Gamma_{\rm s}}{5/3} \right)^{-2.17}  \nonumber \\
& \times \left[ 1 - 0.30 \left(f_{\rm corot}-1 \right) \right],
\end{align}
which we use here to estimate $\langle \gloss \rangle$ and its dependence on binary system parameters. In the expression above, $\gd = \Ma / \Md$ and $\gL \approx 1.2^2 M^2 / (\Md \Ma)$ \citep{1998CoSka..28..101P}.

\subsection{Integrating Coupled Mass and Angular Momentum Loss}

We can integrate the mass loss over an angular momentum interval according to the specific angular momentum loss per unit mass, 
\beq
\Delta m = - \int_{L_{\rm i}}^{L_{\rm f}} l_{\rm loss}^{-1} dL,
\eeq
 where $L_{\rm i}$ and $L_{\rm f}$ are the initial and final angular momenta of the binary. 
If we use the mass-averaged definition of $\langle \gloss \rangle$ and make the approximation that the total binary mass is nearly constant, then we can re-write this as follows,
\beq
\Delta m \approx - \frac{M}{\langle\gamma_{\rm loss}\rangle}  \int_{L_{\rm i}}^{L_{\rm f}} \frac{dL}{L}  =  \frac{M}{\langle\gamma_{\rm loss}\rangle} \ln \left(\frac{L_{\rm i}}{L_{\rm f} } \right).
\eeq
If we again apply the approximation that the binary's total and component masses are nearly constant, we can express this result in terms of initial and final separations, $a_{\rm i}$ and $a_{\rm f}$,
\beq
\Delta m \approx \frac{M}{\langle\gamma_{\rm loss}\rangle} \ln \left[ \left(\frac{a_{\rm i}}{a_{\rm f} } \right)^{1/2} \right].
\eeq
While the approximation of constant component masses is not strictly valid over the course of the binary interaction, we will show in Section \ref{results} that this simplified form is useful in estimating our simulation output. 

We apply this estimate of the total mass loss over an interval in two cases, first to estimate the total mass loss as the binary converges from the Roche limit separation, $a_{\rm RL}$ to the original donor radius, $R_1$,
 \beq\label{dme}
\dme (R_1)  \equiv   \frac{M}{\langle\gamma_{\rm loss}\rangle} \ln \left[ \left(\frac{a_{\rm RL}}{R_1}  \right)^{1/2} \right], 
\eeq
where the Roche limit is defined by the \citet{1983ApJ...268..368E} approximation is 
\beq
{a_{\rm RL} \over R_1} = \frac{0.6q^{-2/3} + \ln ( 1+q^{-1/3} )}{0.49q^{-2/3}},
\eeq
In this expression, we note that our definition of $q=\Ma/\Md$ is the inverse of that of \citet{1983ApJ...268..368E}.

Secondly, our hydrodynamic models are initialized at varying binary separations, which implies varying initial orbital angular momenta. All other things being equal, this would lead to a difference in the total ejecta mass. We compensate for this by integrating the same, averaged specific angular momentum loss over the separations not simulated from the Roche limit to the starting separation of the binary simulation, $a_0$,  and adding this estimate to the numerically computed mass loss over the simulated interval from $a_0$ to $R_1$, which is $\Delta m (R_1)$. Thus, 
\beq\label{dmt}
\dmt(R_1) \equiv \Delta m (R_1)  + \frac{M}{\langle\gamma_{\rm loss}\rangle} \ln \left[ \left(\frac{a_{\rm RL}}{a_0}  \right)^{1/2} \right], 
\eeq
estimates the total mass loss if the simulation were initialized at $a_{\rm RL}$.

\subsection{Results}\label{results}

\begin{figure}[tbp]
\begin{center}
\includegraphics[width=0.49\textwidth]{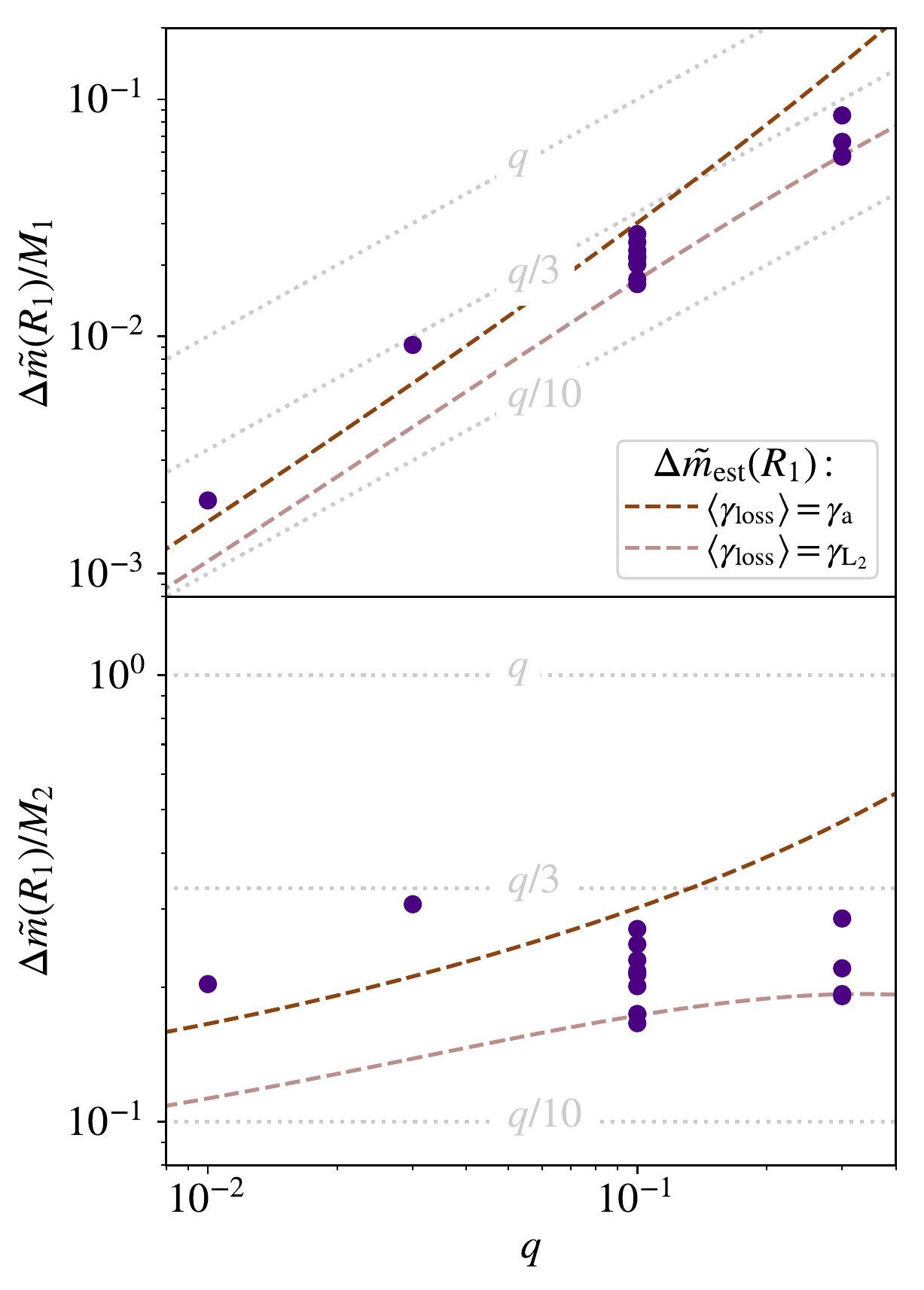}
\caption{Mass removed from the donor star as a function of binary mass ratio, equation \eqref{dmt}. This figure summarizes all of the models of Table \ref{simtable}, and therefore encapsulates other parameter variations in the repeated entries at a given $q$ value.  We observe that $\dmt(R_1)$ is a strong function of $q$, in fact, the lower panel shows that it scales nearly linearly with $q$. For guidance, the dashed lines show $\dme (R_1)$ from equation \eqref{dme}, given different assumptions about the specific angular momentum carried by ejecta, $\langle \gloss \rangle = \gacc$ and $\langle \gloss \rangle = \gL$. We find that $\dmt(R_1)/M_2$ is nearly constant across a wide range in binary mass ratio. }
\label{fig:dmq}
\end{center}
\end{figure}

We begin by examining the primary trend in our simulated model systems, the dependence of $\dmt$, equation \eqref{dmt}, on binary mass ratio, $q$. Figure \ref{fig:dmq} shows the dependence of $\dmt (R_1)$ on $q$, rescaled to the mass of the donor star ($M_1$, top panel) and the accretor star ($M_2$, bottom panel).  Multiple entries at a given $q$ reflect the variations of other parameters in Table \ref{simtable}.  

From Figure \ref{fig:dmq} it is immediately clear that $\dmt(R_1)$ has approximately linear dependence on $q$. As a result, $\dmt(R_1)/M_2$ is nearly a constant across more than an order of magnitude in binary mass ratio. We find that $\dmt(R_1)$ is consistently on the order of 25\% of $M_2$. Earlier analytical and simulation work has been suggestive of a similar scaling and normalization \citep{2017ApJ...835..282M,2018ApJ...868..136M}, but the results presented here represent the first systematic survey of binary parameter space.

\begin{figure}[tbp]
\begin{center}
\includegraphics[width=0.49\textwidth]{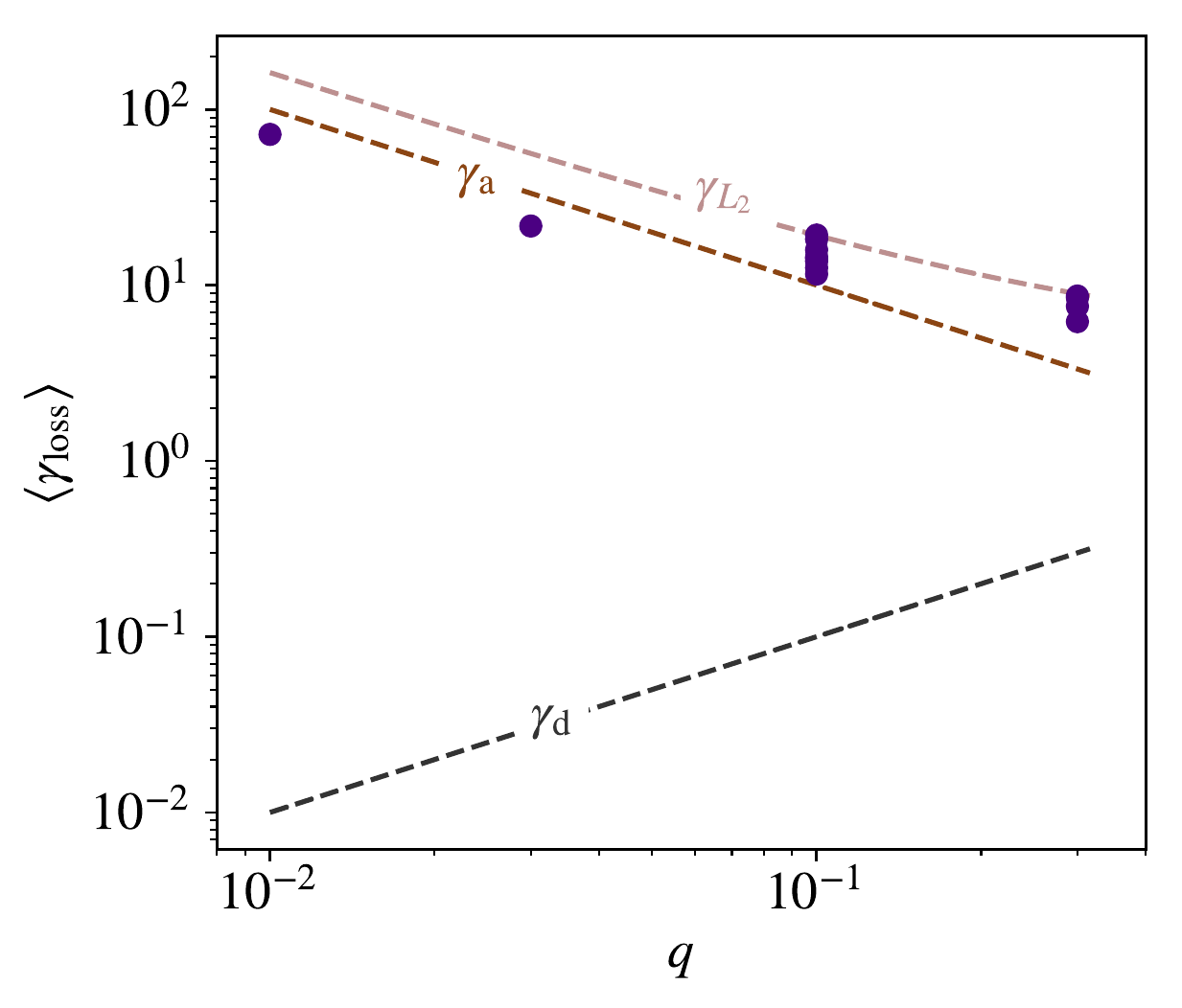}
\caption{Mass averaged dimensionless specific angular momentum with which material is expelled from the binary, $\langle \gloss\rangle$. We compare to reference values of the specific angular momentum of the donor, accretor, and $L_2$ Lagrange point.  After being removed from the donor, material flows toward the accretor and then is lost from the binary with $\langle \gloss\rangle$ more similar to the characteristic values of the accretor, $\gacc$ and $\gL$. These trends allow $\Delta m(R_1)$ to be estimated as a function of binary properties using equation \eqref{dme}. }
\label{fig:gloss}
\end{center}
\end{figure}

The approximate origin of this scaling is best seen by examination of the estimated mass loss, $\dme(R_1)$ of equation \eqref{dme}. To make progress, it is essential to understand the value of $\langle \gloss \rangle$. The numerical values of $\langle \gloss \rangle$ are well approximated by equation \eqref{fit_gamma}. For clarity, we also display the dependence of $\langle \gloss \rangle$ on $q$ in Figure \ref{fig:gloss}. Figure \ref{fig:gloss} shows that the typical value of $\langle \gloss \rangle$ is on the order of the angular momentum of the accretor, $\gacc = \Md/\Ma$, or the $L_2$ Lagrange point $\gL \approx 1.2^2 M^2 / (\Md \Ma)$, and vastly different from the original angular momentum of the material, $\gd$, given its origin in the donor star. Given these values of $\langle \gloss \rangle$, we can use equation \eqref{dme} to estimate
\beq\label{dme_gaL}
 \frac{\dme (R_1) } {\Ma} \sim
    \begin{cases}
      0.3  (1+q), &{\rm for}\  \langle \gloss \rangle = \gacc, \\
      0.2  (1+q)^{-1}, &{\rm for}\ \langle \gloss \rangle = \gL,
    \end{cases}
\eeq
 where, for the simulated systems $\Ma = M_2$. 
We note that the result for $\langle \gloss \rangle = \gL$ is equivalent to $ \dme (R_1) \approx 0.2 \mu$ where $\mu = \Md \Ma / M$ is the binary reduced mass. The numerical coefficients in equation \eqref{dme_gaL} originate from the logarithm term in equation \eqref{dme}; the numerical values given are appropriate for $q=0.15$, and there is a factor of approximately 2 dependence in the range $0.01 \leq q \leq 0.3$.  

In Figure \ref{fig:dmq}, we add lines for $\dme (R_1)$ as computed from the full version of equation \eqref{dme}, but given $ \langle \gloss \rangle = \gacc$ and  $\langle \gloss \rangle = \gL$. These estimates reproduce the approximate magnitude of $\dmt(R_1)$. We find, however, that our numerical results have less slope with $q$. We can attribute this to the variation of $\langle \gloss \rangle$ relative to $\gacc$ and $\gL$. For low $q$, we find $\langle \gloss \rangle < \gacc$, while for higher $q$ we find $\langle \gloss \rangle \gtrsim \gL$ \citep{2019arXiv191205545M}. This slope in $\langle \gloss \rangle$ compensates for the slopes seen in $\dme (R_1)$, and, as a result, $\dmt(R_1)/M_2$ is constant with respect to varying $q$. 

\begin{figure*}[tbp]
\begin{center}
\includegraphics[width=0.99\textwidth]{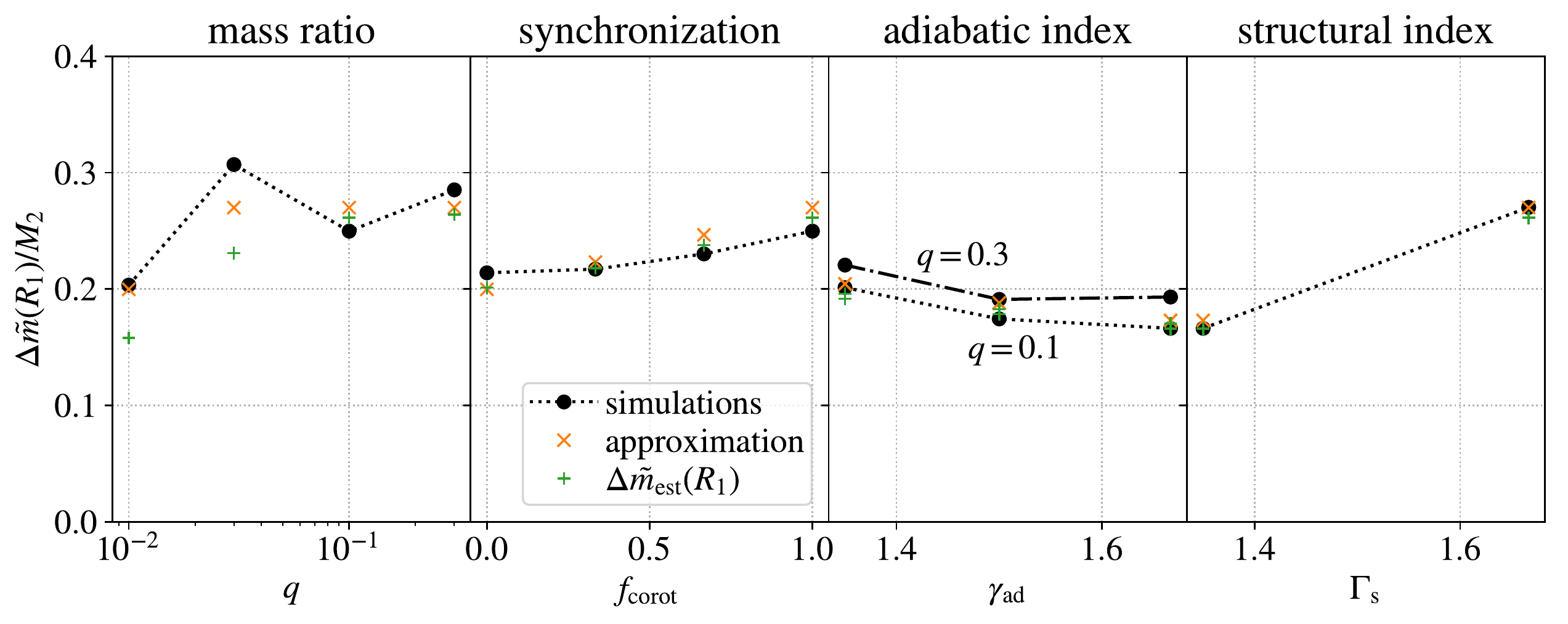}
\caption{Variations in $\dmt(R_1)$ with binary system properties. The panels above isolate varying mass ratio (models E, A, and F), degree of spin-orbit synchronization (models A--D), adiabatic index (models G--I and J--L), and structural index (models M and I).  We note that $\dmt(R_1)/M_2$ is only weakly variant with binary system properties. This implies that the primary dependency is the linear dependence on binary mass ratio, $q$. We compare to two approximating forms, a least-squares fitting formula, equation \eqref{fit_dm}, and $\dme(R_1)$. The fit of $\dme(R_1)$, equation \eqref{dme}, demonstrates that variations seen above reflect differences in the mass-averaged specific angular momentum with which gas is lost from the binary system, $\langle \gloss \rangle$, which is approximated by equation \eqref{fit_gamma}.    }
\label{fig:dmparam}
\end{center}
\end{figure*}

Finally, we consider the dependence of $\dmt (R_1)$ on binary system parameters, as shown in Figure \ref{fig:dmparam}. Overall, we observe relatively mild dependence on binary parameters, $\dmt (R_1)/M_2$ varies by approximately a  factor of two of across binary system parameters. In Figure \ref{fig:dmparam}, we compare to two approximations of $\dmt (R_1)/M_2$. By least-squares fitting, we derive the following approximating formula,
\begin{align}\label{fit_dm}
\frac{\dmt(R_1) }{M_2} \approx& 0.27  \left( \frac{\gamma_{\rm ad}}{5/3} \right)^{-0.79}  \left( \frac{\Gamma_{\rm s}}{5/3} \right)^{2.11}  \nonumber \\
& \times \left[ 1 - 0.26 \left(f_{\rm corot}-1 \right) \right],
\end{align}
in which we note that that power-law dependence on mass ratio was also a fitted parameter, but the best fit slope was consistent with zero when rounded to two decimal places. In Figure \ref{fig:dmparam}, the results of equation \eqref{fit_dm} are labeled ``approximation".  Next, we compare to $\dme (R_1)/M_2$, as computed from equation \eqref{fit_gamma}'s fit to $\langle \gloss \rangle$.  We find that that both of these approximate forms reproduce the main trends of the simulation data, especially with respect to binary synchronization, adiabatic and structural indicies. The dependence on $q$ is somewhat more variable around the predictions than the other quantities (though we note that the lower value and prediction at $q=0.01$ are due to $f_{\rm corot}=0$ rather than $f_{\rm corot}=1$).\footnote{Studying model snapshots, as shown in Appendix A of \citet{2019arXiv191205545M}, we find that this likely has to do with the particularities of how we measure $\Delta m$. The $q=0.03$ exhibits tidal oscillations at $a=R_1$, which raise some mass to $r>R_1$, contributing to our numerical definition of $\Delta m$, without actually contributing to the mass expelled from the binary.}

\section{Circumbinary Torus Formation}\label{sec:torus}

Material that is pulled from the donor expands into the circumbinary environment. In this section, we study the  resulting distribution of stripped material.  

\subsection{Ejecta Binding Energy}

\begin{figure*}[tbp]
\begin{center}
\includegraphics[width=0.99\textwidth]{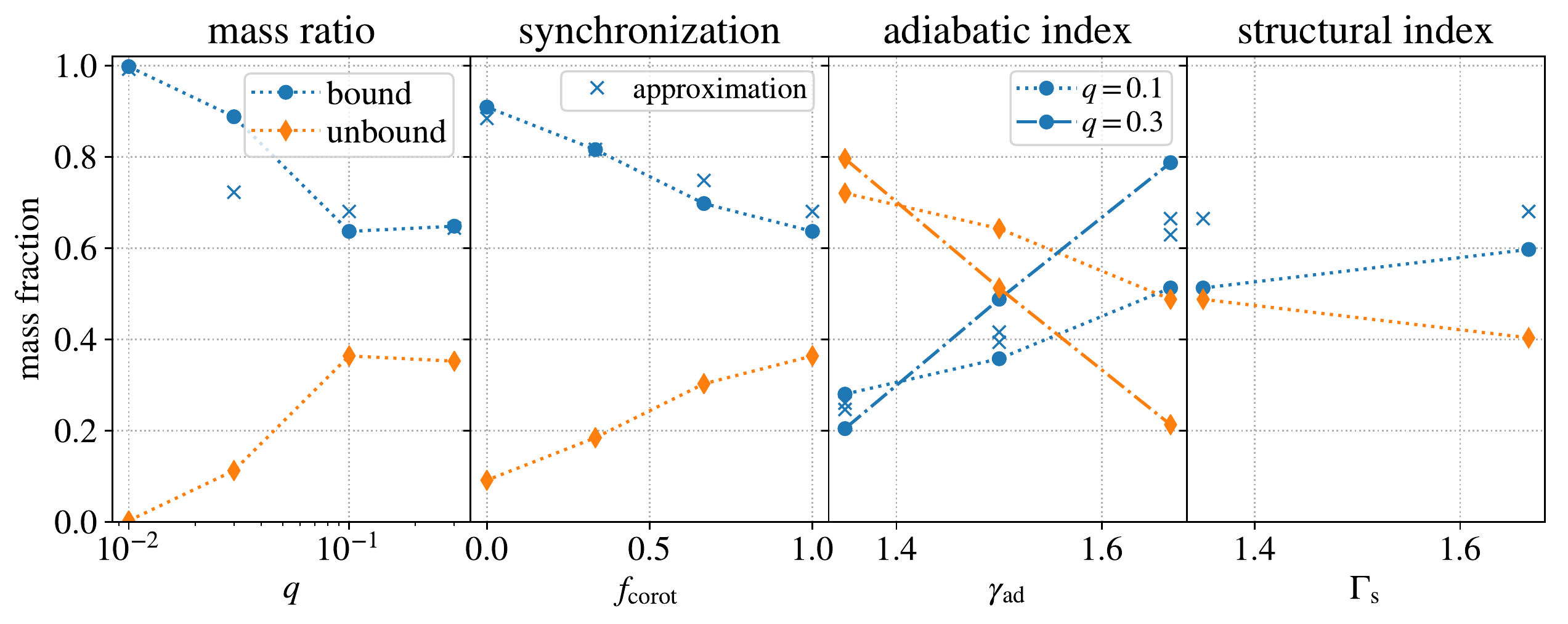}
\caption{Fractions of $\Delta m (R_1)$ that are bound or unbound relative to the binary system, based on Bernoulli parameter in the model snapshot. We find that in each of our models some expelled material remains bound to the binary, while some is unbound. However, the kinematics of the ejecta, and thus these relative proportions are sensitive to the simulation parameters. In particular, higher mass ratios yield a higher fraction of unbound ejecta, and lower adiabatic indices result in stronger gravitational slingshot past the accretor and a higher proportion of unbound ejecta.   }
\label{fig:BE}
\end{center}
\end{figure*}

We evaluate the binding energy of gas expelled into the circumbinary environment relative to the binary as follows. We compute its Bernoulli parameter, $\cal B$, which is the sum of its specific kinetic energy, potential, and enthalpy, as elaborated in Appendix \ref{energyappendix}. We denote material with ${\cal B} >0$ as unbound and material with ${\cal B}<0$ as bound. An important caveat to note here is that these quantities measure an instantaneous representation of the gas, rather than its final state since material is not, in general, on the free streamlines for which the Bernoulli parameter will be constant. Thus, gas that is unbound at one moment may collide with other gas parcels, redistributing its energy and achieving a different division of bound and unbound material \citep{2018ApJ...868..136M}.

Figure \ref{fig:BE} shows the variations in bound and unbound mass fractions of ejecta at the $t=t_1$, when the binary separation equals the original donor radius. Because these are reported as fractions of the ejecta, the bound and unbound mass fractions sum to unity. A key overall observation from Figure \ref{fig:BE} is that a fraction between 20\% to 100\% of the material outflowing from the binary is  bound, and will not likely escape to infinity without further energy or momentum input. \citet{2018ApJ...868..136M} explored this in detail for a single binary model, which had $q=0.3$ and $\gamma_{\rm ad} = \Gamma_{\rm s} = 5/3$ (equivalent to Model F).  They found that primarily bound material was lost early in the trend from Roche lobe overflow to binary coalescence, with higher radial velocity, unbound ejecta being expelled as the separation shrank to be similar to the donor radius \citep{2018ApJ...868..136M}.

Previous hydrodynamic simulations of the common envelope dynamical inspiral phase have similarly found that much of the material flung away from the inspiralling stellar cores remains bound to the binary \citep[for example,][]{2008ApJ...672L..41R,2012ApJ...746...74R,2012ApJ...744...52P,2016ApJ...816L...9O,2016MNRAS.460.3992N,2016MNRAS.462L.121O,2018MNRAS.480.1898C,2019MNRAS.486.5809P}. The bound fractions vary somewhat, but the finding of the majority of gas being bound in hydrodynamic or magnetohydrodynamic simulations with an ideal gas equation of state seems to be universal \citep{2013A&ARv..21...59I}. The addition of other physical processes \citep[section 3.5 of][]{1993PASP..105.1373I} like a more realistic equation of state including ionization and recombination of hydrogen and helium \citep{2015MNRAS.447.2181I,2015MNRAS.450L..39N,2016MNRAS.460.3992N,2016MNRAS.462..362I}, radiation pressure on dust \citep{2018MNRAS.478L..12G}, or mechanical feedback from jets or other outflows \citep{2015ApJ...800..114S,2016RAA....16...99K,2016NewA...47...16S,2017MNRAS.465L..54S,2017MNRAS.471.4839S,2018MNRAS.477.2584S}, have all been discussed as possible mechanisms to eventually unbind circumbinary material.

Returning to the specific context of pre-common envelope orbital decay, Figure \ref{fig:BE} demonstrates that changing binary properties in our models leads to non-negligible differences in the distribution of bound versus unbound mass. While the structure of the donor star has no dramatic affect on the binding energy distribution, other properties, like mass ratio, degree of donor rotation, and gas adiabatic index are all impactful in our results. We find that the following fitting formula broadly approximates our results,
\begin{align}\label{fit_bound}
f_{\rm bound} \approx& 0.68 \left( \frac{q}{0.1} \right)^{-0.05}  \left( \frac{\gamma_{\rm ad}}{5/3} \right)^{4.45}  \left( \frac{\Gamma_{\rm s}}{5/3} \right)^{0.11}  \nonumber \\
& \times \left[ 1 - 0.30 \left(f_{\rm corot}-1 \right) \right],
\end{align}
and we discuss individual dependencies in what follows. 

Upon examination of the simulation output, we find that the distinctions in gas binding energy with binary parameters can be traced to differing outflow kinematics as opposed to thermodynamics. Models with a larger fraction of unbound material have larger radial velocities of outflow relative to the binary center of mass, not larger internal energies (i.e. higher temperatures). The trends observed are thus preserved regardless of the inclusion of gas internal energy or enthalpy in the definition of the binding energy \citep[see, for example, the discussion of][]{2011ApJ...731L..36I}. In particular,  the binary and gas parameters affect gas flow in the crucial interaction and loss region around the accretor. 

Low mass accretors, low $q$, are observed to be less able to impart sufficient impulse to unbind material from the binary than high mass accretors with larger $q$. The higher mass accretors impart both more specific angular momentum relative to that of the binary -- see the fitting formula of equation \eqref{fit_gamma} -- and larger radial velocities to the outflows. This combination of additional kinetic energy and angular momentum imparted per unit fluid mass leads to a higher proportion of unbound mass at larger $q$. Similarly, degree of donor star corotation, $f_{\rm corot}$, affects the flow morphology through the $L_1$ Lagrange point and near the accretor, with these differences reflected in a higher unbound mass fraction for corotating donors. 

Finally, we note that gas adiabatic index, $\gamma_{\rm ad}$ has a dramatic impact on the bound and unbound mass fractions. Interestingly, the greater degree of unbound material when $\gamma_{\rm ad}=1.35$ is not due to the closer-to-isothermal equation of state, but to differences in the gas flow in the immediate vicinity of the accretor. We find that expelled gas' angular momentum is not substantially modified by varying $\gamma_{\rm ad}$ between 1.35 and 5/3, as seen in equation \eqref{fit_gamma} and Figure 5 of \citet{2019arXiv191205545M}. However, the radial velocity of gas leaving the vicinity of the accretor and $L_2$ increases significantly when $\gamma_{\rm ad}$ is lower. This places more material on trajectories that are unbound relative to the binary center of mass. 

We trace the origin of this difference in ejecta radial velocity to pressure gradients near the $L_1$ point and accretor in the simulations. For example, when $\gamma_{\rm ad}=5/3$ and fluid is compressed near the accretor, pressure gradients overwhelm the gravitational force, and the minimal impact parameter relative to the accretor is dictated by the fluid's resistance to compression through this nozzle point. By contrast, the enhanced compressibility of lower $\gamma_{\rm ad}$ cases lead to deeper penetration of the $L_1$ stream toward the accretor, and thus greater momentum and energy transfer.  For a related discussion in slightly different context, see Section 4.1 of \citet{2017ApJ...845..173M}.  Given that the depth of penetration to the accretor appears to be of critical importance, we caution that the particular radial velocity attained may be related to the numerical choice of the accretor softening radius. Astrophysically, this would imply that accretors of differing compactness might yield different outflow properties. 

This discussion highlights two important lessons. On one hand, mass ejection from the binary and its binding energy seems a broad-brush, global property, but our simulations demonstrate that it depends primarily on flow from the donor through the vicinity surrounding the accretor. Without simulating this crucial region, we have little information on the kinematics of the circumbinary material. Secondly, the sensitivities of this flow indicates that unmodeled properties like more realistic equations of state, radiative diffusion and cooling, and magnetohydrodynamical stresses may all play important roles in influencing overall properties beyond the immediate vicinity of the binary and the mass loss region.  

\subsection{Circumbinary Structures}

Gas is lost from the binary preferentially in the plane of the orbit. As it self-intersects in spiral shock waves \citep{2016MNRAS.455.4351P,2016MNRAS.461.2527P,2018ApJ...868..136M}, it also spreads vertically, eventually forming thick, toroidal structures around the binary.  Within these tori, some material reaches a quasi-hydrostatic configuration, while other gas is outflowing. Because these structures are largely axisymmetric about the binary's angular momentum vector, we will examine azimuthally-averaged properties of circumbinary structures in what follows.

\begin{figure}[tbp]
\begin{center}
\includegraphics[width=0.48\textwidth]{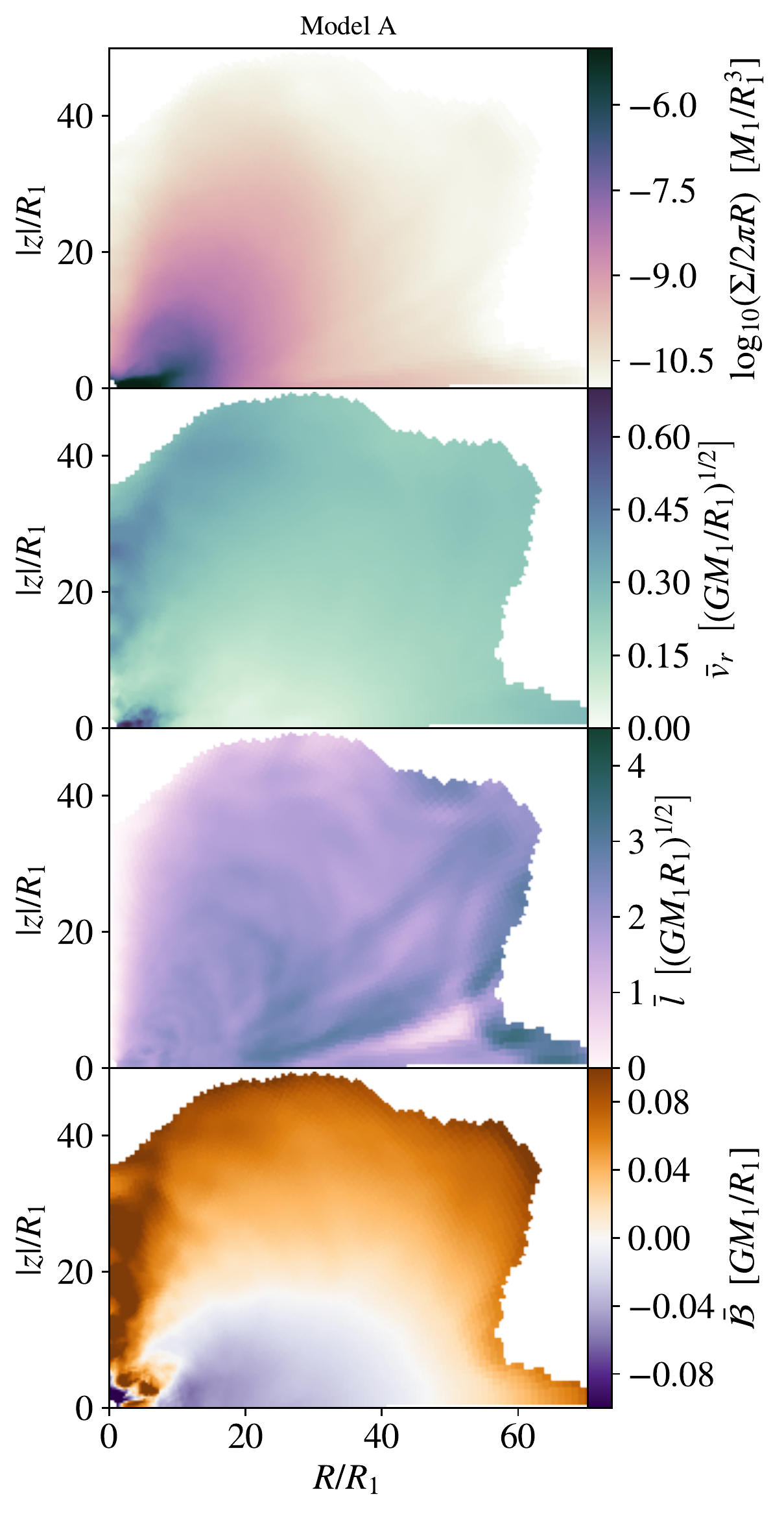}
\caption{Azimuthally-averaged properties of circumbinary structure at $t=t_1$ for model A (properties in Table \ref{simtable}). Structures are azimuthally averaged about the donor star center and above and below the equatorial plane. Gas density shows a thick, toroidal structure with evacuated poles and rapidly decreasing density with distance from the binary. Gas is largely moving with low radial velocity, with the exception of the leading edge of the ejecta and near the pole. The combination $R\bar v_\phi$ is nearly constant throughout the torus, implying that material has nearly constant specific angular momentum throughout. Finally, gas Bernoulli parameter demonstrates that the circumbinary material is composed of a bound torus surrounded by unbound ejecta.  }
\label{fig:az0}
\end{center}
\end{figure}

We compute azimuthally-averaged quantities in the $R$--$z$ plane with origin at the donor star center, like our computational mesh. At $R\gg R_1$ the binary center of mass would be a more ideal choice of origin for the azimuthal average, but the origin of the computational mesh limits our ability to compute quantities in that manner. We also note that the offset between the center of mass and donor center at $t=t_1$ is smaller than $R_1$, so any differences at $R \gg R_1$ are minor. Figure \ref{fig:az0} shows azimuthally-averaged density, $\Sigma / 2\pi R$, and several mass-weighted kinematic properties of simulation model A at $t_1$ -- thus this represents the circumbinary distribution as the binary coalesces.   We denote mass-weighted azimuthally-averaged variables with a bar, such as ${\bar v}_{r}$ for the mass-weighted average radial velocity. 

Figure \ref{fig:az0} shows a toroidal structure of density with relatively evacuated poles, but otherwise very thick scale height, extending to approximately 50 times the original donor star radius. We see that in most regions the radial velocity, $\bar v_r$, is at most 10-20\% of the donor star's escape velocity. Near the poles a low-density region of somewhat more rapid outflow develops as the surrounding torus collimates later ejecta toward the path of least resistance \citep{2018ApJ...868..136M}.

Figure \ref{fig:az0} also shows the specific angular momentum, $\bar l$, relative to the binary center of mass. We observe that $\bar l$ is nearly constant throughout the circumbinary gas. Spiral shocks, launched by the binary's orbital motion, trace their way through the ejecta -- this is seen clearly in Figure 4 of \citet{2016MNRAS.455.4351P} and Figure 1 of \citet{2018ApJ...868..136M}, for example.  These shocks appear to very efficiently redistribute angular momentum within the circumbinary gas, such that the broader distribution of angular momenta with which gas is flung away  from the binary converges to a single value \citep[e.g.][Figures 8 and 11]{2018ApJ...868..136M}.   Interestingly, a torus of constant specific angular momentum is unstable to perturbations, the epicyclic frequency is zero, and the instability has been further demonstrated by linear perturbation theory \citep{1986PThPh..75..251K}. Thus the persistence of the nearly-constant specific angular momentum structures in our simulations must relate to the ongoing energy and momentum injection from the central binary.

Finally, the lower panel of Figure \ref{fig:az0} considers the distribution of azimuthally-averaged Bernoulli parameter within the ejecta. In this example, an extensive equatorial region extending to roughly 50 times the donor's radius consists of material bound to the binary. The leading edge of ejecta and material further from the binary midplane is unbound. The most-unbound material (that with the largest escape velocity) is that in the polar funnel region, as discussed by \citet{2018ApJ...868..136M}. 

We find that the qualitative features are similar across our models, with the most variable aspect being the relative quantities of bound and unbound gas as demonstrated by the Bernoulli parameter, and as discussed in the previous section. Each of the azimuthally-averaged datasets is available as an hdf5 file and equivalent versions of Figure \ref{fig:az0} are also included \citep[online at \url{https://github.com/morganemacleod/PreCEMassLoss} and][]{morgan_macleod_2020_3692070}.

\subsection{Circumbinary Torus Approximation}

Because the presence of circumbinary material in a toroidal configuration is ubiquitous in our models, in this section we consider to what extent these circumbinary structures can be reproduced by simplified, analytic distributions. In general, the circumbinary gas distributions have complexity that is not captured by any analytic model that we considered, as we will discuss in more detail below. In particular, the presence of ongoing outflow and newly-ejected material close to the binary are not captured in the simplified forms we consider. Nonetheless, an analytic approximation may be useful in some applications where the complexity of full simulation output is not desirable or warranted. 

We considered several models and found that one that is useful -- in the sense that it retains an analytic form while capturing some of the gross features of the azimuthally-averaged fluid distribution -- is that of a hydrostatic, barotropic torus of constant angular momentum.  We note here that in isolation, such a torus is unstable to perturbations. However, this model seems to best describe the results of our simulation models with the presence an ongoing source of momentum and energy at the origin.

To model the torus, we consider gas in the torus to be in hydrostatic equilibrium in the cylindrical $R$-$z$ plane surrounding the binary (the binary's angular momentum vector points in the $+z$-direction). Motion in the $\phi$-direction is determined by the condition of constant specific angular momentum, $l_{\rm torus} = R v_\phi$. A barotropic equation of state implies that pressure is a function of density only, we adopt a polytropic form $P = K_{\rm torus} \rho^{\gamma_{\rm torus}}$.  Hydrostatic equilibrium is derived from a steady-state assumption in the gas momentum equation, in which case pressure gradients balance centrifugal forces from gas rotation and gravity. A final parameter, $R_{0, \rm torus}$, describes $R$ at which the torus density vanishes at $z=0$. A derivation and description of this torus model is given in Appendix \ref{torusdef}.  In its simplicity, this model fails to capture model-specific features like inhomogeneities, and ignores any outflows. It is thus best-suited to the cases of mostly-bound circumbinary material. 

We fit the model parameters $K_{\rm torus}$, $\gamma_{\rm torus}$, $l_{\rm torus}$, and $R_{0, \rm torus}$ to each simulation case as follows. First, we determine the extent of the circumbinary material, characterized by $R_{0, \rm torus}$. To do so, we select expelled material from the background gas in the azimuthal average slices by selecting only material which is at least three times the initial ambient density of the outermost zones, and which has entropy less than 0.9 that of the original isentropic background. Finally, we find the average outermost radius of this selected material for a range of angles from $\pi /6 \leq \theta \leq \pi /2$. We assign this angle-averaged extent from the simulation to $R_{0, \rm torus}$. 

We determine the total torus mass and angular momentum, and specify $l_{\rm torus} = L_{\rm torus}/M_{\rm torus}$. We specify that $M_{\rm torus} = \Delta m (R_1)$, where $M_{\rm torus}$ is integrated as in equation \eqref{Mtorus}. 
We fit the parameters of the polytropic equation of state, $K_{\rm torus}$ and $\gamma_{\rm torus}$, by mass-weighted least-squares fitting in the $\bar P$--$\bar \rho$ plane. We apply an additional constraint to this fit that requires that $M_{\rm torus} = \Delta m (R_1)$.

\begin{figure*}[tbp]
\begin{center}
\includegraphics[width=0.34\textwidth]{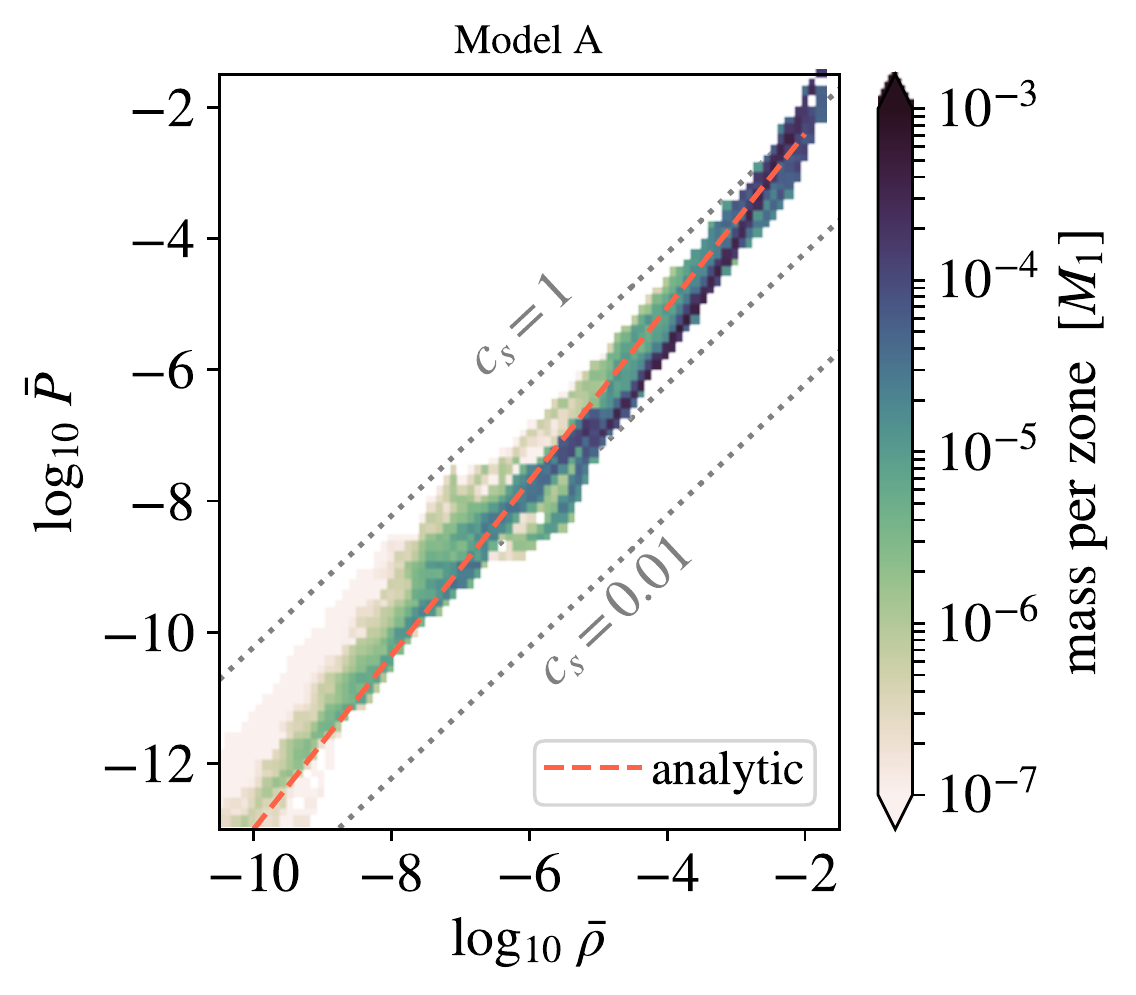}
\includegraphics[width=0.64\textwidth]{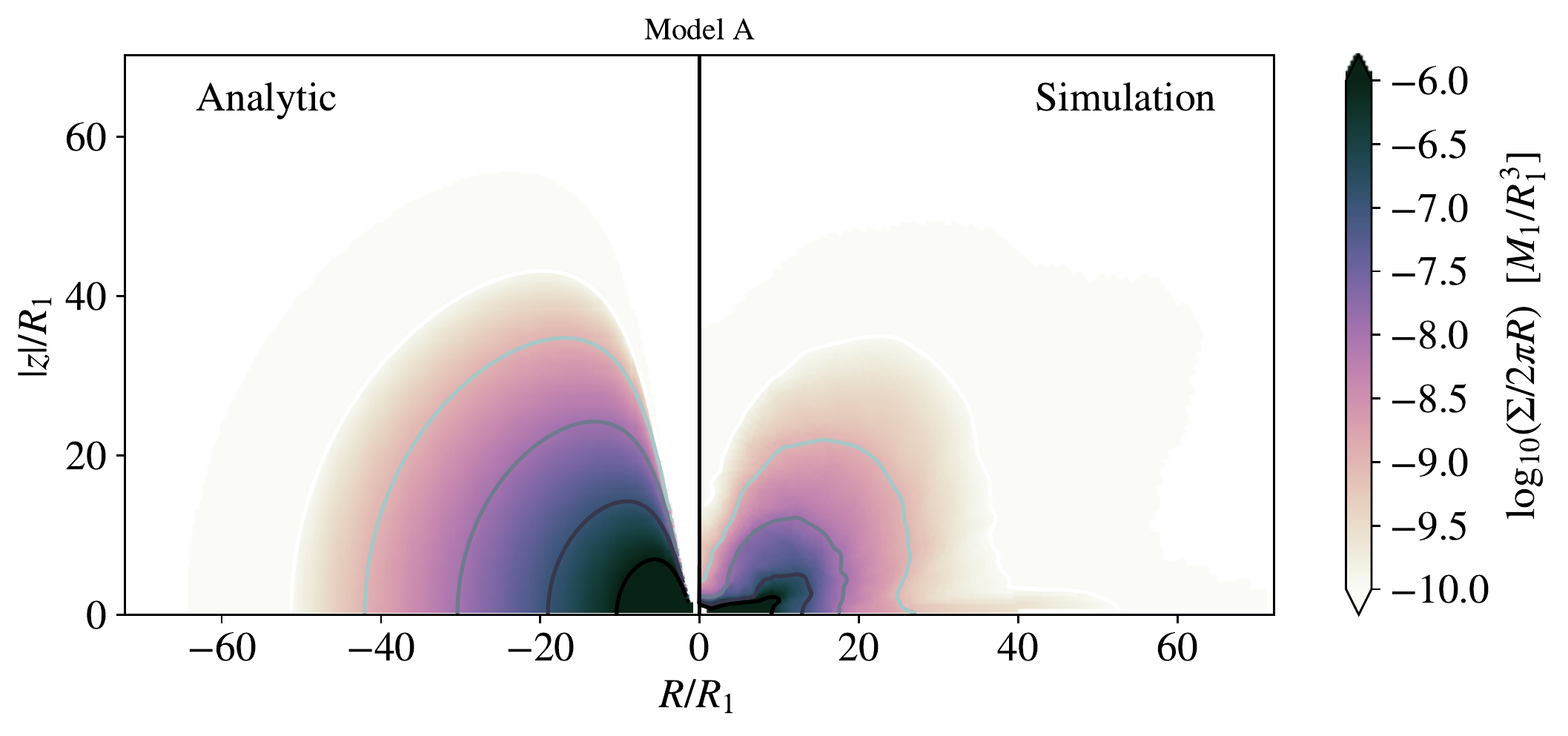}
\caption{Comparison of model A at $t=t_1$ to analytic torus model. The left panel shows  pressure-density phase space of the azimuthally-averaged profile, the polytropic analytic model fits this with a powerlaw. The density profile compares the best-fit quasi-hydrostatic model torus to the simulated circumbinary distribution. The analytic model captures some of the main features of the circumbinary distribution, like the evacuated poles and overall density profile. Features associated with ongoing mass ejection close to the binary, and more distant unbound ejecta are not captured.  }
\label{fig:torus_comp}
\end{center}
\end{figure*}

Figure \ref{fig:torus_comp} \citep[Figure set online at \url{https://github.com/morganemacleod/PreCEMassLoss} and][]{morgan_macleod_2020_3692070}  displays a comparison between the torus model for simulation A and the simulated output.  We compare the distribution of material in the pressure--density plane, along with the powerlaw fit determined by $K_{\rm torus}$ and $\gamma_{\rm torus}$. We then compare the azimuthally averaged density distribution to that of the model. One conclusion to draw from this comparison is that the overall morphology of the two models shares commonalities, like the approximate profile of density -- with higher densities near the equatorial plane and an evacuated polar region. However, significant differences are also present, and largely represent the ways in which the full simulation output is distinct from a hydrostatic torus, in particular with respect to the presence of outflowing material with nonzero radial velocity. 

Table \ref{torustable} displays the derived model torus parameters and their dependence on binary model parameters. We observe that the specific angular momentum expelled in the torus, $l_{\rm torus}$, and the torus extent, $R_{\rm 0,torus}$, both grow with mass ratio.  Otherwise, the binary parameter dependence of torus model parameters is not pronounced. A second conclusion is that the range of values for $\gamma_{\rm torus} \sim 1.2$ to 1.4,  is relatively compact, and the values exhibit little parameter dependence. The fact that in most cases $\gamma_{\rm torus} \ll \gamma_{\rm ad}$ when $\gamma_{\rm ad}=5/3$ implies the significant role that shock heating plays in increasing the entropy of torus gas as it expands and density decreases. Spiral shocks launched by the binary's motion clearly mediate both the angular momentum content and thermodynamics of this circumbinary material \citep{2016MNRAS.455.4351P,2016MNRAS.461.2527P,2017ApJ...850...59P,2017MNRAS.471.3200M}. \citet{2016MNRAS.455.4351P} has shown that a key factor is the velocity dispersion with which material leaves the binary since this plays a role in determining the relative velocity of internal shocks. 

In this discussion of gas thermodynamics, it is important to note that our hydrodynamic simulations include no radiative diffusion or cooling. Once heat content is established through gas dynamical processes such as those we discuss, it can also be lost through radiation \citep{2016MNRAS.455.4351P}. Processes of molecule and dust formation and their associated opacity effects are of crucial importance here and impart scale dependence on astrophysical binary outcomes through the microphysics \citep[e.g. as discussed by][]{2017MNRAS.471.3200M}. \citet{2016MNRAS.461.2527P} has usefully classified outcomes for bound circumbinary tori in hierarchies of timescale: comparing rates of cooling to heating, for example. Thus, the thermodynamic conditions implied by our simplified results correspond directly only to the limit where radiative diffusion is unimportant on the dynamical timescale throughout the model.

\begin{table}
\begin{center}
\begin{tabular}{ccccc}
Name & $l_{\rm torus}$ & $R_{0,\rm torus}$ & $K_{\rm torus}$ & $\gamma_{\rm torus}$ \\
\hline
A & 1.26 & 63.88 & 1.74 & 1.32 \\
B &1.36 & 59.97 & 1.65 & 1.32 \\
C & 1.44 & 58.43 & 2.40 & 1.35 \\
D & 1.46 & 29.74 & 4.33 & 1.41 \\
\hline
N & 0.72 & 9.95 & 7.67 & 1.38 \\
E & 0.75 & 27.10 & 0.44 & 1.20 \\
F & 1.50 & 89.45 & 0.79 & 1.28 \\
\hline
G & 1.49 & 50.00 & 0.71 & 1.27 \\
H & 1.74 & 48.39 & 1.69 & 1.33 \\
I & 1.81 & 41.86 & 1.23 & 1.31 \\
\hline
J & 1.77 & 89.65 & 0.63 & 1.27 \\
K & 2.05 & 96.70 & 2.19 & 1.35 \\
L & 1.99 & 96.19 & 1.81 & 1.33 \\
\hline
M & 1.12 & 9.80 & 0.79 & 1.31 \\
\hline
\end{tabular}
\caption{Best-fit analytic model torus parameters. The (constant) specific angular momentum of torus material is $l_{\rm torus}$ and its outer radial extent is $R_{0,\rm torus}$. The models' barotropic equation of state is described by $P = K_{\rm torus} \rho^{\gamma_{\rm torus}}$.  }
\label{torustable}
\end{center}
\end{table}

\section{Implications of Pre-Coalescence Mass Loss}\label{sec:implications}

In this section, we discuss the implications of a more-informed connection between binary properties and mass ejection during the pre-coalescence phase. We suggest that this connection represents a step toward connecting binary systems to the transients they produce and toward new predictive models of common envelope phase outcomes \citep[e.g.][]{2016MNRAS.460.3992N,2019MNRAS.490.2550I}.

\subsection{Common Envelope Observational Appearance}

Circumbinary material generated by pre-coalescence mass loss may have at least two implications for the observational appearance of binaries undergoing this evolution. First, circumbinary material is likely to obscure and modify the appearance of the binary \citep{2014ApJ...788...22P}, and second, the presence of significant circumbinary medium may lead to shock heating of later ejecta and changes in associated transient light curves \citep{2016MNRAS.455.4351P,2016MNRAS.461.2527P,2017MNRAS.471.3200M}. 

As material populates the circumbinary environment, it may become optically thick and advance the photosphere beyond the stellar surfaces \citep{2014ApJ...788...22P,2017ApJ...850...59P}. In this case, the appearance of the binary will change as the bolometric luminosity remains similar, but the effective temperature decreases while the radiating surface area grows \citep{2011A&A...528A.114T}. Varying opacities as a function of density and temperature are likely to be crucial in modulating the appearance of the system during this stage. For example, when hydrogen goes from ionized to atomic with decreasing temperature, the opacity can decrease by several orders of magnitude, leading to an optical photosphere at a characteristic temperature for hydrogen recombination of several thousand kelvin \citep{2013Sci...339..433I,2016MNRAS.455.4351P,2016MNRAS.461.2527P,2017ApJ...835..282M}. Similarly, in colder gas ($\lesssim 10^3$K) dust may form, dramatically increasing the opacity at optical wavelengths \citep{2016MNRAS.455.4351P,2017MNRAS.471.3200M,2018MNRAS.478L..12G,2019MNRAS.489.3334I,2020arXiv200306151I}. In these cases, the binary would be expected to exhibit a low effective temperature and very extended emitting surface, perhaps tens or a hundred times the original radii of the stars. 

A second important aspect of circumbinary material is its influence on potential later ejecta \citep[e.g.][]{2017MNRAS.471.3200M}. In the context of supernovae explosions and similar transients, the influence of circumstellar material on light curve peak magnitudes and morphologies is well documented, as shock heating increases the thermal energy of ejecta at radii close to the transient photosphere \citep[e.g.][]{2013MNRAS.433..838P}. In the specific context of common envelope phases and coalescing binary systems, \citet{2017MNRAS.471.3200M} have developed a semi-analytic model based on this scenario in which an outburst meets a surrounding circumbinary distribution. This model captures some aspects of the behavior observed in our simulations and those of \citet{2016MNRAS.455.4351P,2016MNRAS.461.2527P}. On the basis of our simulations, a difference that may be important is that emitted material from the binary appears to self-interact in series of spiral shock waves rather than in a single outburst. 

Quantifying the kinematics, spatial distribution, and mass of circumbinary material as well as the later ejecta are of crucial importance to developing semi-analytic models of transient appearance, and in those regards the results of this work represent a step toward those goals.

\subsection{Common Envelope Simulations}

The stage of progressive Roche lobe overflow is slow compared to the rapid orbital tightening that occurs during the dynamical phase of a common envelope episode \citep{2018ApJ...863....5M}. This is due to the rapidly rising forces that arise from the plunge of the accretor within the envelope of the donor \citep{2018ApJ...863....5M,2019MNRAS.490.3727C}. As a result, many global common envelope simulations are initialized with the accretor already at the surface of the donor \citep[e.g.][]{2016ApJ...816L...9O,2016MNRAS.462L.121O,2018MNRAS.480.1898C,2019MNRAS.486.5809P}, in order to devote computational resources to the phase of most rapid and dramatic interaction. Our results show that this approximation is potentially problematic, because it neglects the presence of substantial quantities circumbinary material still present. 

In particular, the leading edge of common envelope ejecta will encounter circumbinary material rather than escaping freely to infinity. This may affect the distribution of bound or unbound material, the ejected gas' kinematics, and thermodynamics. Each of these properties influence the outcome of a common envelope phase because the orbiting binary cores must clear and unbind any surrounding circumbinary material before their orbit fully stabilizes \citep{2016MNRAS.462..362I,2016MNRAS.461..486K}. 

However, encapsulated in our results, we also propose a simplified circumbinary torus model   that captures some of the overall features of the circumbinary material  \citep[Appendix \ref{torusdef}, and available with examples via version 1.1 of our {\tt RLOF} python software,][]{RLOF1.1}.\footnote{online at \url{https://github.com/morganemacleod/RLOF/releases/tag/v.1.1}} Initializing a common envelope simulation with initial separation equal to the donor radius but with this circumbinary distribution is simple and would be more appropriate than with no circumbinary material.

\section{Summary and Conclusion}\label{sec:conclusion}

We have studied hydrodynamic simulations of coalescing binary systems undergoing unstable Roche lobe overflow from an extended donor star to a more-compact accretor. We analyzed how mass is expelled away from the vicinity of the binary, with particular focus on its total mass and kinematics.  
Some key conclusions of this work are as follows. 
\begin{enumerate}

\item Across a range of binary mass ratios, the expelled mass pre-coalescence -- from Roche lobe overflow until the separation decays to the donor's original radius -- is approximately 25\% of the accretor's mass, as shown in Figures \ref{fig:dmq} and \ref{fig:dmparam}, and approximated by equations \eqref{dme} and \eqref{fit_dm}. 

\item This ejected material is composed of a range of both unbound and bound material. These kinematics and sensitive in particular to the gas equation of state and binary mass ratio -- both of which affect the flow of gas through the nozzle near $L_1$ and away from the binary near $L_2$ (Figure \ref{fig:BE}). The bound fraction is estimated by  equation \eqref{fit_bound}. 

\item At the time of coalescence, previously expelled material is assembled into a roughly toroidal circumbinary distribution. Torus material has nearly-constant specific angular momentum and, especially in cases of mostly-bound material, low radial velocities (Figure \ref{fig:az0}). We show that a hydrostatic equilibrium torus with polytropic equation of state and constant specific angular momentum provides an analytic approximation for the circumbinary distribution (Figure \ref{fig:torus_comp} and Appendix \ref{torusdef}). 

\item The presence of  circumbinary material at the onset of common envelope phases has implications for their observable signatures, as circumbinary material enshrouds the inner binary \citep{2014ApJ...788...22P} and spiral shocks heat circumbinary material \citep{2016MNRAS.455.4351P,2016MNRAS.461.2527P,2017MNRAS.471.3200M}. The simple relationship between circumbinary properties and binary mass ratio will prove useful in initializing future common envelope hydrodynamic simulations from contact between the binary components. 
 
\end{enumerate}

There are several crucial caveats to bear in mind regarding the use and interpretation of the results and fitting formulae presented in this paper. While we model the hydrodynamics of the binary interaction and mass loss from the binary system, the physics of this interaction is, by necessity, simplified. It is informative to consider which parameterizations induce differences in the results. For example, we have found that the pre-common envelope mass loss is relatively independent of binary parameters when scaled to the mass of the accretor, $\dmt (R_1) / M_2$ (Figure \ref{fig:dmparam}). On the other hand, the kinematics of this outflow do appear to depend, with some sensitivity, on the parameters of the interaction (Figure \ref{fig:BE}). The details of the gas flow through the nozzle at $L_1$ and then away from the binary, appear especially important. In this vein, changing the parameters of the equation of state affects the kinematics of ejecta and, in turn, the distribution of specific binding energy within the circumbinary gas. The lesson to take from this sensitivity is not the particular numerical result, but that the properties of outflow depend on the conditions near the accretor. 

The range of possible conditions near the accretor include numerous physical effects that go beyond the scope of what we have considered in this paper. The general equation of state will have a compressibility that depends on the density--temperature regime, and may therefore approximate the respective limits of our survey of $\gamma_{\rm ad}$ in different portions of the same binary system. The accretors themselves have properties beyond their gravitational influence, which is all that we model in this paper. The respective compactness (whether the accretor is a main sequence star or compact object) and magnetic field of the accretor are examples of properties that are influential in observed binary systems, and may also be critical in these scenarios \citep{2015ApJ...800..114S,2016RAA....16...99K,2016NewA...47...16S,2017MNRAS.465L..54S,2017MNRAS.471.4839S,2018MNRAS.477.2584S}. 

Finally, we note that the circumbinary distribution of material that results during the binary coalescence highlights the significance of radiative cooling that parts of this material will undergo. With the large range of gas densities, there will always be regions in which cooling is important on the dynamical timescale of the flow. Because at least a portion of the torus is bound to the binary, the outcome of this radiative cooling may have dynamical significance in the ongoing interaction traced by spiral shocks through the circumbinary material. \citet{2016MNRAS.461.2527P} have insightfully characterized bound circumbinary torii in terms of the ratio of cooling timescale to mass ejection timescale, which can play an important role in determining whether the torus is thick, as in our simulations without radiative losses, or thinner and more equatorially concentrated due to loss of pressure support \citep{2016MNRAS.461.2527P}. Because these ratios of timescales depend on binary properties, like the system mass and radius, understanding these dependencies is an important goal for future work. 

Despite these cautions on the range of lessons that can be drawn from the present results, we emphasize that our findings to date represent a significant step toward the goal of uniting binary systems with the observable transients they produce through their interactions. In particular, because we find that the pre-common envelope mass ejecta depends primarily -- and relatively robustly -- on the mass of the engulfed accretor object, we can associate the quantity of pre-common envelope circumbinary material with the properties of the subsumed object. This is valuable because, even when a pre-outburst ``progenitor" is detected -- as in the case of the transients M31 LRN 2015 \citep{2015ApJ...805L..18W,2015A&A...578L..10K,2017ApJ...835..282M} and M101 OT2015-1 \citep{2017ApJ...834..107B} -- most of the emission comes from the donor star and we typically do not have the benefit of also observing lower-mass companion. Thus, linking properties that may eventually be inferred from photometric observations, such as the ejecta mass, to the underlying binary may become especially useful.

\acknowledgements{ We thank M. Mapelli and S. Toonen for helpful discussions and insights. M.M. is grateful to E.C. Ostriker and J. M. Stone for their extensive contributions to the development of the hydrodynamic methods on which this work is based. We thank an anonymous referee for thoughtful and detailed feedback. 

We acknowledge support for this work provided by NASA through Einstein Postdoctoral Fellowship grant number PF6-170169 awarded by the Chandra X-ray Center, which is operated by the Smithsonian Astrophysical Observatory for NASA under contract NAS8-03060. 
This material is based upon work supported by the National Science Foundation under Grant No. 1909203. 
Resources supporting this work were provided by the NASA High-End Computing (HEC) Program through the NASA Advanced Supercomputing (NAS) Division at Ames Research Center. }

\software{IPython \citep{PER-GRA:2007}; SciPy \citep{jones_scipy_2001};  NumPy \citep{van2011numpy};  matplotlib \citep{Hunter:2007}; Astropy \citep{2013A&A...558A..33A}; Athena++ (Version 1.1.1, Stone, J.M., \url{https://github.com/PrincetonUniversity/athena-public-version}); RLOF, version 1.1 \citep{RLOF1.1} }

\clearpage
\bibliographystyle{aasjournal}

\begin{thebibliography}{}
\expandafter\ifx\csname natexlab\endcsname\relax\def\natexlab#1{#1}\fi

\bibitem[{jon(2001)}]{jones_scipy_2001}
 2001, {SciPy}: Open source scientific tools for Python, ,

\bibitem[{{Astropy Collaboration} {et~al.}(2013){Astropy Collaboration},
  {Robitaille}, {Tollerud}, {Greenfield}, {Droettboom}, {Bray}, {Aldcroft},
  {Davis}, {Ginsburg}, {Price-Whelan}, {Kerzendorf}, {Conley}, {Crighton},
  {Barbary}, {Muna}, {Ferguson}, {Grollier}, {Parikh}, {Nair}, {Unther},
  {Deil}, {Woillez}, {Conseil}, {Kramer}, {Turner}, {Singer}, {Fox}, {Weaver},
  {Zabalza}, {Edwards}, {Azalee Bostroem}, {Burke}, {Casey}, {Crawford},
  {Dencheva}, {Ely}, {Jenness}, {Labrie}, {Lim}, {Pierfederici}, {Pontzen},
  {Ptak}, {Refsdal}, {Servillat}, \& {Streicher}}]{2013A&A...558A..33A}
{Astropy Collaboration}, {Robitaille}, T.~P., {Tollerud}, E.~J., {et~al.} 2013,
  \aap, 558, A33

\bibitem[{{Blagorodnova} {et~al.}(2017){Blagorodnova}, {Kotak}, {Polshaw},
  {Kasliwal}, {Cao}, {Cody}, {Doran}, {Elias-Rosa}, {Fraser}, {Fremling},
  {Gonzalez-Fernand ez}, {Harmanen}, {Jencson}, {Kankare}, {Kudritzki},
  {Kulkarni}, {Magnier}, {Manulis}, {Masci}, {Mattila}, {Nugent}, {Ochner},
  {Pastorello}, {Reynolds}, {Smith}, {Sollerman}, {Taddia}, {Terreran},
  {Tomasella}, {Turatto}, {Vreeswijk}, {Wozniak}, \&
  {Zaggia}}]{2017ApJ...834..107B}
{Blagorodnova}, N., {Kotak}, R., {Polshaw}, J., {et~al.} 2017, \apj, 834, 107

\bibitem[{{Chamandy} {et~al.}(2019){Chamandy}, {Blackman}, {Frank},
  {Carroll-Nellenback}, {Zou}, \& {Tu}}]{2019MNRAS.490.3727C}
{Chamandy}, L., {Blackman}, E.~G., {Frank}, A., {et~al.} 2019, \mnras, 490,
  3727

\bibitem[{{Chamandy} {et~al.}(2018){Chamandy}, {Frank}, {Blackman},
  {Carroll-Nellenback}, {Liu}, {Tu}, {Nordhaus}, {Chen}, \&
  {Peng}}]{2018MNRAS.480.1898C}
{Chamandy}, L., {Frank}, A., {Blackman}, E.~G., {et~al.} 2018, \mnras, 480,
  1898

\bibitem[{{Eggleton}(1983)}]{1983ApJ...268..368E}
{Eggleton}, P.~P. 1983, \apj, 268, 368

\bibitem[{{Galaviz} {et~al.}(2017){Galaviz}, {De Marco}, {Passy}, {Staff}, \&
  {Iaconi}}]{2017ApJS..229...36G}
{Galaviz}, P., {De Marco}, O., {Passy}, J.-C., {Staff}, J.~E., \& {Iaconi}, R.
  2017, \apjs, 229, 36

\bibitem[{{Glanz} \& {Perets}(2018)}]{2018MNRAS.478L..12G}
{Glanz}, H., \& {Perets}, H.~B. 2018, \mnras, 478, L12

\bibitem[{{Hernquist} \& {Katz}(1989)}]{1989ApJS...70..419H}
{Hernquist}, L., \& {Katz}, N. 1989, \apjs, 70, 419

\bibitem[{{Howitt} {et~al.}(2020){Howitt}, {Stevenson}, {Vigna-G{\'o}mez},
  {Justham}, {Ivanova}, {Woods}, {Neijssel}, \& {Mandel}}]{2020MNRAS.492.3229H}
{Howitt}, G., {Stevenson}, S., {Vigna-G{\'o}mez}, A.~r., {et~al.} 2020, \mnras,
  492, 3229

\bibitem[{Hunter(2007)}]{Hunter:2007}
Hunter, J.~D. 2007, Computing In Science \& Engineering, 9, 90

\bibitem[{{Iaconi} \& {De Marco}(2019)}]{2019MNRAS.490.2550I}
{Iaconi}, R., \& {De Marco}, O. 2019, \mnras, 490, 2550


\bibitem[{{Iaconi} {et~al.}(2017){Iaconi}, {Reichardt}, {Staff}, {De Marco},
  {Passy}, {Price}, {Wurster}, \& {Herwig}}]{2017MNRAS.464.4028I}
{Iaconi}, R., {Reichardt}, T., {Staff}, J., {et~al.} 2017, \mnras, 464, 4028

\bibitem[Iaconi et al.(2019)]{2019MNRAS.489.3334I} Iaconi, R., Maeda, K., De Marco, O., et al.\ 2019, \mnras, 489, 3334

\bibitem[Iaconi et al.(2020)]{2020arXiv200306151I} Iaconi, R., Maeda, K., Nozawa, T., et al.\ 2020, arXiv e-prints, arXiv:2003.06151

\bibitem[{{Iben} \& {Livio}(1993)}]{1993PASP..105.1373I}
{Iben}, Icko, J., \& {Livio}, M. 1993, \pasp, 105, 1373

\bibitem[{{Ivanova} \& {Chaichenets}(2011)}]{2011ApJ...731L..36I}
{Ivanova}, N., \& {Chaichenets}, S. 2011, \apjl, 731, L36

\bibitem[{{Ivanova} {et~al.}(2013{\natexlab{a}}){Ivanova}, {Justham}, {Avendano
  Nandez}, \& {Lombardi}}]{2013Sci...339..433I}
{Ivanova}, N., {Justham}, S., {Avendano Nandez}, J.~L., \& {Lombardi}, J.~C.
  2013{\natexlab{a}}, Science, 339, 433

\bibitem[{{Ivanova} {et~al.}(2015){Ivanova}, {Justham}, \&
  {Podsiadlowski}}]{2015MNRAS.447.2181I}
{Ivanova}, N., {Justham}, S., \& {Podsiadlowski}, P. 2015, \mnras, 447, 2181

\bibitem[{{Ivanova} \& {Nandez}(2016)}]{2016MNRAS.462..362I}
{Ivanova}, N., \& {Nandez}, J.~L.~A. 2016, \mnras, 462, 362

\bibitem[{{Ivanova} {et~al.}(2013{\natexlab{b}}){Ivanova}, {Justham}, {Chen},
  {De Marco}, {Fryer}, {Gaburov}, {Ge}, {Glebbeek}, {Han}, {Li}, {Lu}, {Marsh},
  {Podsiadlowski}, {Potter}, {Soker}, {Taam}, {Tauris}, {van den Heuvel}, \&
  {Webbink}}]{2013A&ARv..21...59I}
{Ivanova}, N., {Justham}, S., {Chen}, X., {et~al.} 2013{\natexlab{b}}, \aapr,
  21, 59

\bibitem[{{Jencson} {et~al.}(2019){Jencson}, {Kasliwal}, {Adams}, {Bond}, {De},
  {Johansson}, {Karambelkar}, {Lau}, {Tinyanont}, {Ryder}, {Cody}, {Masci},
  {Bally}, {Blagorodnova}, {Castell{\'o}n}, {Fremling}, {Gehrz}, {Helou},
  {Kilpatrick}, {Milne}, {Morrell}, {Perley}, {Phillips}, {Smith}, {van Dyk},
  \& {Williams}}]{2019ApJ...886...40J}
{Jencson}, J.~E., {Kasliwal}, M.~M., {Adams}, S.~M., {et~al.} 2019, \apj, 886,
  40

\bibitem[{{Kashi} \& {Soker}(2016)}]{2016RAA....16...99K}
{Kashi}, A., \& {Soker}, N. 2016, Research in Astronomy and Astrophysics, 16,
  99
  
 \bibitem[Kojima(1986)]{1986PThPh..75..251K} Kojima, Y.\ 1986, Progress of Theoretical Physics, 75, 251

\bibitem[{{Kurtenkov} {et~al.}(2015){Kurtenkov}, {Pessev}, {Tomov},
  {Barsukova}, {Fabrika}, {Vida}, {Hornoch}, {Ovcharov}, {Goranskij}, {Valeev},
  {Moln{\'a}r}, {S{\'a}rneczky}, {Kostov}, {Nedialkov}, {Valenti}, {Geier},
  {Wiersema}, {Henze}, {Shafter}, {Mu{\~n}oz Dimitrova}, {Popov}, \&
  {Stritzinger}}]{2015A&A...578L..10K}
{Kurtenkov}, A.~A., {Pessev}, P., {Tomov}, T., {et~al.} 2015, \aap, 578, L10

\bibitem[{{Kuruwita} {et~al.}(2016){Kuruwita}, {Staff}, \& {De
  Marco}}]{2016MNRAS.461..486K}
{Kuruwita}, R.~L., {Staff}, J., \& {De Marco}, O. 2016, \mnras, 461, 486

\bibitem[{MacLeod(2020{\natexlab{a}})}]{morgan_macleod_2020_3692070}
MacLeod, M. 2020{\natexlab{a}}, morganemacleod/PreCEMassLoss: v1.0, v.v1.0,
  Zenodo, doi:10.5281/zenodo.3692070

\bibitem[{MacLeod(2020{\natexlab{b}})}]{RLOF1.0}
---. 2020{\natexlab{b}}, morganemacleod/RLOF: v1.0, v.v1.0,  Zenodo,
  doi:10.5281/zenodo.3663006

\bibitem[{MacLeod(2020{\natexlab{c}})}]{RLOF1.1}
---. 2020{\natexlab{c}}, morganemacleod/RLOF: v1.1, v.v.1.1,  Zenodo,
  doi:10.5281/zenodo.3690127

\bibitem[{{MacLeod} \& {Loeb}(2019)}]{2019arXiv191205545M}
{MacLeod}, M., \& {Loeb}, A. 2019, arXiv e-prints, arXiv:1912.05545

\bibitem[{{MacLeod} {et~al.}(2017){MacLeod}, {Macias}, {Ramirez-Ruiz},
  {Grindlay}, {Batta}, \& {Montes}}]{2017ApJ...835..282M}
{MacLeod}, M., {Macias}, P., {Ramirez-Ruiz}, E., {et~al.} 2017, \apj, 835, 282

\bibitem[{{MacLeod} {et~al.}(2018{\natexlab{a}}){MacLeod}, {Ostriker}, \&
  {Stone}}]{2018ApJ...868..136M}
{MacLeod}, M., {Ostriker}, E.~C., \& {Stone}, J.~M. 2018{\natexlab{a}}, \apj,
  868, 136

\bibitem[{{MacLeod} {et~al.}(2018{\natexlab{b}}){MacLeod}, {Ostriker}, \&
  {Stone}}]{2018ApJ...863....5M}
---. 2018{\natexlab{b}}, \apj, 863, 5

\bibitem[{{MacLeod} {et~al.}(2019){MacLeod}, {Vick}, {Lai}, \&
  {Stone}}]{2019ApJ...877...28M}
{MacLeod}, M., {Vick}, M., {Lai}, D., \& {Stone}, J.~M. 2019, \apj, 877, 28

\bibitem[{{Metzger} \& {Pejcha}(2017)}]{2017MNRAS.471.3200M}
{Metzger}, B.~D., \& {Pejcha}, O. 2017, \mnras, 471, 3200

\bibitem[{{Murguia-Berthier} {et~al.}(2017){Murguia-Berthier}, {MacLeod},
  {Ramirez-Ruiz}, {Antoni}, \& {Macias}}]{2017ApJ...845..173M}
{Murguia-Berthier}, A., {MacLeod}, M., {Ramirez-Ruiz}, E., {Antoni}, A., \&
  {Macias}, P. 2017, \apj, 845, 173

\bibitem[{{Nandez} \& {Ivanova}(2016)}]{2016MNRAS.460.3992N}
{Nandez}, J.~L.~A., \& {Ivanova}, N. 2016, \mnras, 460, 3992

\bibitem[{{Nandez} {et~al.}(2014){Nandez}, {Ivanova}, \&
  {Lombardi}}]{2014ApJ...786...39N}
{Nandez}, J.~L.~A., {Ivanova}, N., \& {Lombardi}, J.~C., J. 2014, \apj, 786, 39

\bibitem[{{Nandez} {et~al.}(2015){Nandez}, {Ivanova}, \&
  {Lombardi}}]{2015MNRAS.450L..39N}
{Nandez}, J.~L.~A., {Ivanova}, N., \& {Lombardi}, J.~C.~J. 2015, \mnras, 450,
  L39

\bibitem[{{Ohlmann} {et~al.}(2016{\natexlab{a}}){Ohlmann}, {R{\"o}pke},
  {Pakmor}, \& {Springel}}]{2016ApJ...816L...9O}
{Ohlmann}, S.~T., {R{\"o}pke}, F.~K., {Pakmor}, R., \& {Springel}, V.
  2016{\natexlab{a}}, \apjl, 816, L9

\bibitem[{{Ohlmann} {et~al.}(2016{\natexlab{b}}){Ohlmann}, {R{\"o}pke},
  {Pakmor}, {Springel}, \& {M{\"u}ller}}]{2016MNRAS.462L.121O}
{Ohlmann}, S.~T., {R{\"o}pke}, F.~K., {Pakmor}, R., {Springel}, V., \&
  {M{\"u}ller}, E. 2016{\natexlab{b}}, \mnras, 462, L121

\bibitem[{{Pan} {et~al.}(2013){Pan}, {Patnaude}, \&
  {Loeb}}]{2013MNRAS.433..838P}
{Pan}, T., {Patnaude}, D., \& {Loeb}, A. 2013, \mnras, 433, 838

\bibitem[{{Passy} {et~al.}(2012){Passy}, {De Marco}, {Fryer}, {Herwig},
  {Diehl}, {Oishi}, {Mac Low}, {Bryan}, \& {Rockefeller}}]{2012ApJ...744...52P}
{Passy}, J.-C., {De Marco}, O., {Fryer}, C.~L., {et~al.} 2012, \apj, 744, 52

\bibitem[{{Pastorello} {et~al.}(2019){Pastorello}, {Mason}, {Taubenberger},
  {Fraser}, {Cortini}, {Tomasella}, {Botticella}, {Elias-Rosa}, {Kotak},
  {Smartt}, {Benetti}, {Cappellaro}, {Turatto}, {Tartaglia}, {Djorgovski},
  {Drake}, {Berton}, {Briganti}, {Brimacombe}, {Bufano}, {Cai}, {Chen},
  {Christensen}, {Ciabattari}, {Congiu}, {Dimai}, {Inserra}, {Kankare},
  {Magill}, {Maguire}, {Martinelli}, {Morales-Garoffolo}, {Ochner}, {Pignata},
  {Reguitti}, {Sollerman}, {Spiro}, {Terreran}, \&
  {Wright}}]{2019A&A...630A..75P}
{Pastorello}, A., {Mason}, E., {Taubenberger}, S., {et~al.} 2019, \aap, 630,
  A75

\bibitem[{{Pejcha}(2014)}]{2014ApJ...788...22P}
{Pejcha}, O. 2014, \apj, 788, 22

\bibitem[{{Pejcha} {et~al.}(2016{\natexlab{a}}){Pejcha}, {Metzger}, \&
  {Tomida}}]{2016MNRAS.461.2527P}
{Pejcha}, O., {Metzger}, B.~D., \& {Tomida}, K. 2016{\natexlab{a}}, \mnras,
  461, 2527

\bibitem[{{Pejcha} {et~al.}(2016{\natexlab{b}}){Pejcha}, {Metzger}, \&
  {Tomida}}]{2016MNRAS.455.4351P}
---. 2016{\natexlab{b}}, \mnras, 455, 4351

\bibitem[{{Pejcha} {et~al.}(2017){Pejcha}, {Metzger}, {Tyles}, \&
  {Tomida}}]{2017ApJ...850...59P}
{Pejcha}, O., {Metzger}, B.~D., {Tyles}, J.~G., \& {Tomida}, K. 2017, \apj,
  850, 59

\bibitem[{P\'erez \& Granger(2007)}]{PER-GRA:2007}
P\'erez, F., \& Granger, B.~E. 2007, Computing in Science and Engineering, 9,
  21

\bibitem[{{Pribulla}(1998)}]{1998CoSka..28..101P}
{Pribulla}, T. 1998, Contributions of the Astronomical Observatory Skalnate
  Pleso, 28, 101

\bibitem[{{Prust} \& {Chang}(2019)}]{2019MNRAS.486.5809P}
{Prust}, L.~J., \& {Chang}, P. 2019, \mnras, 486, 5809

\bibitem[{{Ricker} \& {Taam}(2008)}]{2008ApJ...672L..41R}
{Ricker}, P.~M., \& {Taam}, R.~E. 2008, \apjl, 672, L41

\bibitem[{{Ricker} \& {Taam}(2012)}]{2012ApJ...746...74R}
---. 2012, \apj, 746, 74

\bibitem[{{Shiber} {et~al.}(2017){Shiber}, {Kashi}, \&
  {Soker}}]{2017MNRAS.465L..54S}
{Shiber}, S., {Kashi}, A., \& {Soker}, N. 2017, \mnras, 465, L54

\bibitem[{{Shiber} \& {Soker}(2018)}]{2018MNRAS.477.2584S}
{Shiber}, S., \& {Soker}, N. 2018, \mnras, 477, 2584

\bibitem[{{Soker}(2015)}]{2015ApJ...800..114S}
{Soker}, N. 2015, \apj, 800, 114

\bibitem[{{Soker}(2016)}]{2016NewA...47...16S}
---. 2016, \na, 47, 16

\bibitem[{{Soker}(2017)}]{2017MNRAS.471.4839S}
---. 2017, \mnras, 471, 4839

\bibitem[{{Soker} \& {Tylenda}(2003)}]{2003ApJ...582L.105S}
{Soker}, N., \& {Tylenda}, R. 2003, \apjl, 582, L105

\bibitem[{{Soker} \& {Tylenda}(2006)}]{2006MNRAS.373..733S}
---. 2006, \mnras, 373, 733

\bibitem[{{Tylenda} {et~al.}(2011){Tylenda}, {Hajduk}, {Kami{\'n}ski},
  {Udalski}, {Soszy{\'n}ski}, {Szyma{\'n}ski}, {Kubiak}, {Pietrzy{\'n}ski},
  {Poleski}, {Wyrzykowski}, \& {Ulaczyk}}]{2011A&A...528A.114T}
{Tylenda}, R., {Hajduk}, M., {Kami{\'n}ski}, T., {et~al.} 2011, \aap, 528, A114

\bibitem[{Van Der~Walt {et~al.}(2011)Van Der~Walt, Colbert, \&
  Varoquaux}]{van2011numpy}
Van Der~Walt, S., Colbert, S.~C., \& Varoquaux, G. 2011, Computing in Science
  \& Engineering, 13, 22

\bibitem[{{Williams} {et~al.}(2015){Williams}, {Darnley}, {Bode}, \&
  {Steele}}]{2015ApJ...805L..18W}
{Williams}, S.~C., {Darnley}, M.~J., {Bode}, M.~F., \& {Steele}, I.~A. 2015,
  \apjl, 805, L18

\end{thebibliography}

\appendix

\section{Gas Energetics}\label{energyappendix}

We use gas' Bernoulli parameter as a measure of its instantaneous specific binding energy to the binary system. The Bernoulli parameter relates to gas' total specific energy through $\mathcal{B} =  \varepsilon_{\rm tot} + P/\rho$, where the addition of $P/\rho$ represents the potential energy associated with gas pressure's ability to do work along a free streamline.  Bernoulli parameter is, therefore, constant along free streamlines. In  the complex and self-intersecting flow around the binary, we caution that energy may be redistributed. Thus the instantaneous state at the time of binary coalescence does not necessarily represent the final energetics of a fluid parcel. 

The Bernoulli parameter is defined as 
\beq\label{bernoulli}
\mathcal{B} = \Phi + h + \varepsilon_{\rm k} 
\eeq
where $\Phi$ is the binary potential,
 $\varepsilon_{\rm k}$ is the specific kinetic energy, and $h$ is the specific enthalpy.   
Our model's binary potential is 
\beq
\Phi = \Phi_1 + \Phi_2 + \Phi_{\rm sg},
\eeq
which represents the gravitational potentials of the donor and accretor core particles and the self-gravitational potential of the gas, respectively. The potential of the accretor is approximated by a softened point mass \citep[][equation A2]{1989ApJS...70..419H}. The  self-gravitational potential is approximated by the undisturbed potential of the donor star (as described in detail in \citet{2018ApJ...863....5M}, section 3.2.2).  Because $\Delta m (R_1) / M_1 \ll 1$, this crude approximation is reasonable for the evolutionary stages that we consider.  
The specific kinetic energy is 
\beq
\varepsilon_{\rm k} = {1\over 2}  v_{\rm com}^2,
\eeq
where $v_{\rm com}$ denotes the inertial-frame velocity relative to the system center of mass. The specific enthalpy is 
\beq
h = \frac{\gamma P}{(\gamma-1)\rho},
\eeq
where $P$ is the pressure and $\rho$ is the density.

\section{Analytical Circumbinary Torus}\label{torusappendix}

\subsection{Polytropic, Constant Angular Momentum Torus}\label{torusdef}

We describe a torus with constant specific angular momentum and polytropic equation of state about a central, gravitating mass. We ignore the self-gravity of the torus material and define pressure gradients such that the torus  is in hydrostatic equilibrium in the $R$--$z$ plane.  We will work in cylindrical ($R,z$) coordinates, where $r=\sqrt{R^2 + z^2}$, and origin at the center of mass. For simplicity, our notation in this subsection is self-consistent, but not consistent with our description of the simulation models, which have a different coordinate origin at the donor center for the computational mesh.

From the gas momentum equation, the equilibrium condition for a torus reads
\beq
\frac{\nabla P }{\rho} - {\bf g} + \frac{l^2}{R^3}  {\bf \hat R} = 0
\eeq
Where ${\bf g} = - GM {\bf \hat r} /r^2$ is the gravitational acceleration due to a central mass $M$ and $l=Rv_\phi$ is the specific angular momentum. Pressure gradients balance the effective acceleration set by the combination of centrifugal and gravitational forces.   We can integrate this expression given an equation of state and choice of several parameters. 

We adopt a polytropic equation of state, $P=K\rho^\gamma$, and assume that $l$=constant. Under those conditions, the $R$-component of the equation reads
\beq
\frac{1}{\rho} \frac{dP}{d R} = - \frac{GM}{R^2} + \frac{l^2}{R^3}
\eeq
We then make the substitution $dP = K \gamma \rho^{\gamma-1} d\rho$, and integrate both sides. We have
\beq
K \gamma \int_0^{\rho} \rho'^{\gamma-2} d\rho' = \int_{R_0}^{R} \left( -\frac{GM}{R'^2} + \frac{l^2}{R'^3}\right) dR',
\eeq
where we define $R_0$ to be the radius in the midplane at which $\rho \rightarrow 0$. Thus
\beq
\frac{K \gamma }{\gamma-1} \rho^{\gamma -1} = \left( \frac{GM}{R} - \frac{l^2}{2R^2}\right) - \left( \frac{GM}{R_0} - \frac{l^2}{2R_0^2}\right),
\eeq
and, therefore, 
\beq
\rho(R,0) = \left( \frac{\gamma-1}{K\gamma} \left[ \left( \frac{GM}{R} - \frac{l^2}{2R^2}\right) - \left( \frac{GM}{R_0} - \frac{l^2}{2R_0^2}\right) \right] \right)^{\frac{1}{\gamma-1}}.
\eeq
The $z$-component of the equilibrium condition reads
\beq
\frac{1}{\rho} \frac{dP}{d z} = - \frac{GMz}{(R^2+z^2)^{3/2}} 
\eeq
We again substitute $dP = K \gamma \rho^{\gamma-1} d\rho$ and integrate vertically,
\beq
K \gamma \int_{\rho(R,0)}^{\rho} \rho'^{\gamma-2} d\rho' = \int_0^z - \frac{GMz'}{(R^2+z'^2)^{3/2}} dz'.
\eeq
After integration, we have
\beq
\frac{K \gamma }{\gamma-1} \left[ \rho^{\gamma -1} - \rho(R,0)^{\gamma -1} \right] = GM \left[ \frac{1}{r} - \frac{1}{R} \right] 
\eeq
or, solving for $\rho$, 
\beq
\rho(R,z) = \left(  \rho(R,0)^{\gamma -1}  +  \frac{\gamma-1}{K\gamma} GM \left[ \frac{1}{r} - \frac{1}{R} \right] \right)^{\frac{1}{\gamma-1}}.
\eeq
Combing these expressions, we find, 
\beq
\rho(R,z) = \left[ \frac{\gamma-1}{K\gamma} \left( \frac{GM}{r} - \frac{l^2}{2R^2} - \frac{GM}{R_0} + \frac{l^2}{2R_0^2} \right) \right]^{\frac{1}{\gamma-1}},
\eeq
which, along with the equation of state and $l$ defines the density, pressure, and velocity fields. 

 A torus of constant specific angular momentum, though in equilibrium, is unstable to perturbations \citep{1986PThPh..75..251K}. One stability criterion is that $dl / dR > 0$, in which case the epicyclic frequency, 
\beq
\kappa^2 = \frac{2\Omega }{ R} \frac{d}{dR}\left(R^2 \Omega\right) =  \frac{2\Omega }{ R}\frac{d l}{dR} ,
\eeq
is real and nonzero.   We therefore caution that the persistence of these structures in our hydrodynamic models is apparently related to the ongoing source of momentum and energy as more material is expelled from the central binary. The long term behavior of these tori, with this source term removed, is unknown.

\subsection{Implementation in {\tt RLOF} Software}

We update the   {\tt RLOF}  software \citep{2019arXiv191205545M,RLOF1.0},  to compute the circumbinary torus corresponding to a given model orbit integration. The new version release is version 1.1 \citep[online at \url{https://github.com/morganemacleod/RLOF/releases/tag/v.1.1}]{RLOF1.1}. In this way, one can model the evolution of unstable Roche lobe overflow, and obtain a simplified estimate of the circumbinary distribution of gas expelled from the binary.  

We define a torus model in terms of the free parameters $l_{\rm torus}$, $R_{0,{\rm torus}}$, $K_{\rm torus}$, and $\gamma_{\rm torus}$, where these parameters correspond to their non-subscripted versions in \ref{torusdef}. 
The total torus mass is, therefore, 
\beq\label{Mtorus}
M_{\rm torus} = 2\pi \int \int R \rho(R,z) \ dR \ dz
\eeq
which scales with the polytropic constant as $M_{\rm torus} \propto K_{\rm torus}^{-1/(\gamma_{\rm torus}-1)}$. 

We define the model parameters as follows. 
The specific angular momentum is based on the orbital integration as
\beq
l_{\rm torus} = \frac{\Delta L }{\Delta m},
\eeq
the binary's change in angular momentum over the its change in mass, which thus assumes that all mass and angular momentum lost from the binary are supplied to the torus. 
We approximate $R_{0,{\rm torus}}$ on the basis of our simulation models as
\beq
R_{0,{\rm torus}} \approx 200 R_1 q_0^{0.64},
\eeq
where $q_0$ is the binary mass ratio prior to Roche lobe overflow. 
We set $\gamma_{\rm torus} = 4/3$ for all models, noting that most simulations have similar values in Table \ref{torustable}. Finally, we set $K_{\rm torus}$ such that $M_{\rm torus} = \Delta m$. In addition to this simulation-specific implementation, we also include the option to manually generate a torus with any parameter combination.

\end{document}